\documentclass{emulateapj}
\usepackage{apjfonts}
\usepackage{hyperref}
\usepackage{graphicx}

\newcommand{\beq}{\begin{equation}}
\newcommand{\eeq}{\end{equation}}

\newcommand\alp{\mbox{$\alpha$}}

\newcommand\solar{\mbox{$_{\normalsize\odot}$}}

\newcommand{\AmS}{{\protect\the\textfont2
  A\kern-.1667em\lower.5ex\hbox{M}\kern-.125emS}}
\newcommand{\lsim}{\ \raise
-2.truept\hbox{\rlap{\hbox{$\sim$}}\raise5.truept\hbox{$<$}\ }}
\newcommand{\gsim}{\ \raise
-2.truept\hbox{\rlap{\hbox{$\sim$}}\raise5.truept\hbox{$>$}\ }}
\newcommand{\simsim}{\ \raise
-2.truept\hbox{\rlap{\hbox{$\sim$}}\raise5.truept\hbox{$\sim$}\ }}

\def\deg{\hbox{$^\circ$}}
\def\Q{\ifmmode\mathcal{Q}\else$\mathcal{Q}$\fi}

\slugcomment{Accepted for publication in the Astrophysical Journal}

\righthead{Hierarchical Stellar Structures in the Local Group Dwarf  Galaxy NGC~6822}
\lefthead{Gouliermis, et al.}

\shorttitle{Hierarchical Stellar Structures in the Local Group Dwarf  Galaxy NGC~6822}
\shortauthors{Gouliermis, et al.}

\begin{document}
 
\title{Hierarchical Stellar Structures in the Local Group Dwarf  Galaxy NGC~6822}



\author{Dimitrios A. Gouliermis}
\affil{Max Planck Institute for Astronomy, 
         K\"onigstuhl 17, 69117 Heidelberg, Germany; dgoulier@mpia-hd.mpg.de}

\author{Stefan Schmeja, Ralf S. Klessen\altaffilmark{1}}
\affil{Zentrum f\"ur Astronomie der Universit\"at Heidelberg, 
         Institut f\"ur Theoretische Astrophysik, \\ Albert-Ueberle-Str.~2, 
         69120 Heidelberg, Germany;\\  sschmeja@ita.uni-heidelberg.de; rklessen@ita.uni-heidelberg.de}



\altaffiltext{1}{Kavli Institute for Particle Astrophysics and 
Cosmology, Stanford University, Menlo Park, CA 94025, USA}

\author{W.~J.~G.~de~Blok}
\affil{University of Cape Town,
         Private Bag X3, Rondebosch 7701, South Africa;\\ edeblok@ast.uct.ac.za}

\and

\author{Fabian Walter}
\affil{Max Planck Institute for Astronomy, 
         K\"onigstuhl 17, 69117 Heidelberg, Germany;\\ walter@mpia-hd.mpg.de}

\begin{abstract} 

We present a comprehensive study of the star cluster population 
and the hierarchical structure in the clustering of blue stars with 
ages \lsim~500~Myr in the Local Group dwarf irregular 
galaxy NGC~6822. Our observational material comprises the most complete 
optical stellar catalog of the galaxy from imaging 
with the Suprime-Cam at the 8.2-m {\sc Subaru} Telescope. 
We identify 47 distinct star clusters with the application of 
the {\sl nearest-neighbor density} method to this catalog 
for a detection threshold of 3$\sigma$ above the average 
stellar density. The size distribution of the detected clusters can be 
very well approximated by a Gaussian with a peak at 
$\sim$~68~pc. The total stellar masses of the clusters are estimated 
by extrapolating the cumulative observed stellar mass function
of all clusters to be in the range $10^3$~-~$10^{4}$~M{\solar}.
Their number distribution is fitted very well by a power-law 
with index $\alpha \sim 1.5 \pm 0.7$, which is consistent with 
the cluster mass functions of other 
Local Group galaxies and the {\sl cluster initial mass function}. 

The application of the {\sl nearest-neighbor density} 
method for various density thresholds, other than 3$\sigma$, 
enabled the identification of stellar concentrations in various 
length-scales, in addition to the detected star clusters of the galaxy.  
The stellar density maps constructed with this technique 
provide a direct proof of hierarchically structured stellar 
concentrations in NGC~6822, in the sense that smaller dense 
stellar concentrations are located {\sl inside} larger and looser
ones. We illustrate this hierarchy by the so-called {\sl dendrogram}, or 
structure tree of the detected stellar structures,
which demonstrates that most of the detected structures split up into 
several substructures over at least three levels. 
We quantify the hierarchy of these structures with the use of the {\sl minimum 
spanning tree} method. We find that structures detected at 1, 2, and 3$\sigma$ 
density thresholds are hierarchically constructed with a fractal 
dimension of $D \approx 1.8$. Some of the larger stellar concentrations, 
particularly in the northern part of the 
central star-forming portion of the galaxy, coincide with IR-bright complexes 
previously identified with {\sl Spitzer} and associated with high column 
density neutral gas, indicating  structures that currently form stars. The 
morphological hierarchy in stellar clustering, which we observe in NGC~6822 
resembles that of the turbulent interstellar matter, suggesting that turbulence on 
pc- and kpc-scales has been probably the major agent that regulated 
clustered star formation in NGC~6822. 

\end{abstract}


\section{Introduction}

Star formation in galaxies is observed on a wide range of scales, from 
stellar complexes and aggregates over OB associations to compact 
embedded clusters, which again often show substructure. These systems 
are not distinct, independent entities but rather appear to form a continuous 
hierarchy of structures over all these scales \citep[e.g.][]{efremov+elmegreen98, 
elmegreen00}. The interstellar matter (ISM) also shows a hierarchical structure 
from the largest giant molecular clouds down to individual clumps and cores.
The complex hierarchical structure of the ISM is believed to be shaped by supersonic
turbulence \citep[e.g.,][]{ballesteros-paredes07}. The scaling relations observed in 
molecular clouds \citep{larson81} can be explained by the effect of turbulence,
where energy is injected at very large scales and cascades down to the smallest
scales, creating eddies and leading to a clumpy, filamentary structure on all scales. 
These structures appear self-similar and are therefore often described as fractal.
Turbulence is believed to play a major role in star formation by creating density
enhancements that become gravitationally unstable and collapse to form stars.
The spatial distribution of young stars and star clusters probably reflects this
process.

There are only few cases, where the formation of stellar structures in star-forming 
galaxies with resolvable stars can be studied in the complete extent of the galaxy. NGC~6822 is one of these 
rare galaxies, because it is far enough from us for its whole extent to be fairly covered in 
small fields-of-view, but also close enough for its bright stellar content to be sufficiently resolved.
NGC~6822 (IC~4895) is a Local Group dwarf irregular of type Ir~IV-V 
\citep{vandenbergh00}. It is the closest dwarf irregular apart from the 
Magellanic Clouds (MCs), located at a distance of $490\pm 40$ kpc from us \citep{mateo98}, 
and belongs to an extended cloud of irregulars named the ``Local Group Cloud''. 
Due to its close distance NGC~6822 appears quite extended on the sky; its optical angular 
diameter is over a quarter of a degree, while its H{\sc i} disk measures close to a
degree \citep{deblok00}. NGC~6822 is relatively near the Galactic plane ($b = -18.4\deg$) and 
therefore suffers from significant foreground extinction from the Milky Way (MW). \cite{massey07} 
using multi-band imaging of NGC~6822 within the {\sl Survey of Local Group Galaxies Currently 
Forming Stars}, derived a total interstellar reddening of $E(B-V) = 0.25 \pm 0.02$, with 
$E(B-V) = 0.22$ being Galactic. Other recent optical studies of the stellar content of the galaxy
\citep{deblok03, battinelli03} show that NGC~6822 has an extended stellar distribution,
with the young blue stars following the distribution of the H{\sc i} disk \citep{komiyama03}, while
the old and intermediate-age stellar population is found significantly more extended than the 
{\sc Hi} disk \citep{deblok06}. 

\begin{figure}[t!]
\centerline{\includegraphics[clip=true,width=0.9\columnwidth]{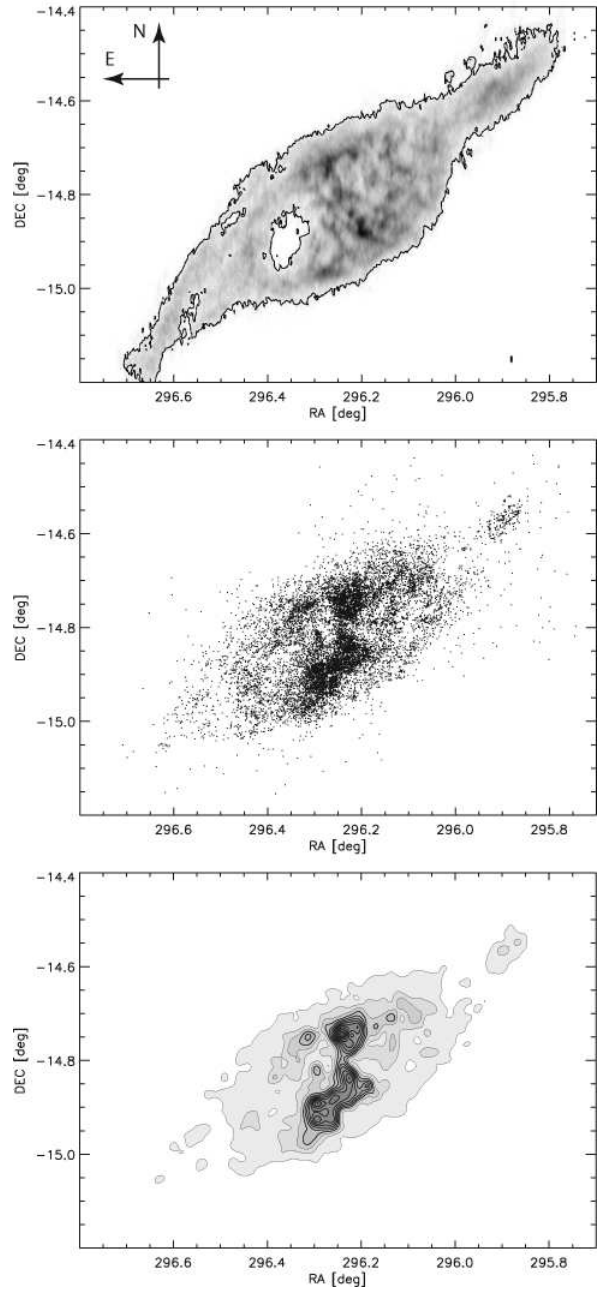}}
\caption{Three views of NGC~6822. {\sl Top:} Integrated ATCA {\sc Hi} column density or 
zeroth-moment map of NGC~6822. The contour indicates the edge of the  {\sc Hi} disk of the 
galaxy at $5 \times 10^{20}$~cm$^{-2}$. The grayscale levels run up to $4.3\times 
10^{21}$~cm$^{-2}$ (black), which is the maximum column density occurring in this map. 
The beam of 42\farcs4~$\times$~12\farcs0 is indicated in the bottom right. This column density 
map is adapted from dBW06, for additional representations see \cite{weldrake03}.
{\sl Middle:} Spatial distribution of the blue stars with $B-R \leq 0.5$ on the sky. 
This distribution highlights the concentration of the blue stars with
ages \lsim~500~Myr in the inner parts of the galaxy, and the general extent 
of these stars in a disk-like structure, similar to the {\sc Hi} disk of the galaxy. {\sl Bottom:} 
Number density map of the blue stars shown in the middle panel, measured in 64\arcsec~$\times$~64\arcsec\ boxes. 
The iso-density contour levels run from the mean background density in steps of 
1$\sigma$, $\sigma$ being the standard deviation of the background density. Isopleths 
of density $\geq 5\sigma$ are plotted with thicker lines.\label{f:maps}}
\end{figure}

The stellar and gas components of NGC~6822 
could not explain the shape of the high-resolution rotation curve obtained with the ATCA, except of the very
inner regions \citep{weldrake03}, and therefore it is considered
to be very dark-matter-dominated. NGC~6822 is a metal-poor, relatively gas-rich galaxy with ISM metal 
abundance of about 0.2~$Z_{\odot}$, i.e. the total fraction of metals is $Z\!\simeq\!0.004$ \citep{skillman89}, and total 
H{\sc i} mass  of $\sim 1.3  \times 10^8\ M_{\odot}$ \citep{deblok00}. Its star formation 
rate (SFR), based on H$\alpha$ and far-IR fluxes, is found to be around  
$\sim 0.06\ M_{\odot}{\rm yr}^{-1}$ \citep{mateo98, israel96} or $\sim 
0.01\ M_{\odot}{\rm yr}^{-1}$ \citep{hunter04}. Evidence for increased star formation 
between 75 and 100 Myr ago is found by \cite{hodge80}, 
while \citet{gallart96a} found that the star formation in NGC~6822 increased by a factor of 2
to 6 between 100 and 200 Myr ago. Moreover, HST imaging showed that the recent SFR is 
spatially variable in the central parts of NGC 6822 \citep{wyder01}. All these findings are 
consistent with the mostly constant but stochastic recent star formation histories often derived 
for other dwarf irregular galaxies \citep[see, e.g.,][]{weisz08}.  

NGC~6822, like most dwarf irregulars, provides an ideal laboratory for the study of galaxy 
evolution, and in particular at early stages, since its low metallicity and rich gas content suggest that 
NGC~6822 is still in an early stage of its conversion from gas into stars. Moreover, its relatively 
simple structure, without dominant spiral arms or bulge, makes the study of the various physical 
processes related to star formation relatively straightforward. The scope of the 
present study is the comprehensive understanding of the structural behavior of star formation 
during the period of the last $\sim$~500~Myr over the whole extent of NGC~6822. 
Specifically, we investigate the cluster population of the galaxy in terms of the size, mass and 
structural characteristics of the detected stellar systems. We also study the clustering behavior 
of stars and its relation to hierarchically structured stellar concentrations within the whole extend of 
NGC~6822. 


A comprehensive study of the star complexes, i.e., large-scale star-forming regions with sizes 
of the order of few hundreds pc, was recently presented by \cite{karampelas09}. 
These authors provide a complete catalog of star complexes in NGC~6822, their positions and sizes,  
based on the data collected within the {\sl Survey of Local Group Galaxies Currently 
Forming Stars} \citep[SLGGCFS;][]{massey06, massey07}.
This study is limited to the brightest stars of the galaxy with 
$B$~\lsim~22.5 due to the detection limit of the SLGGCFS and only to its central part, 
which is covered by the survey. In our investigation here we extend the study 
of the spatial distribution of young stars and hierarchical star formation in NGC~6822 to the 
smallest detectable structures, and to the complete extent of the galaxy, including the potentially 
tidal systems, with the use of original 
photometric material that comprises a stellar sample 6 times larger than that used by \cite{karampelas09}. 
Our catalog extends to fainter magnitudes by 
$B \simeq 2.5$~mag on the main sequence, and covers the whole area of the H{\sc i} disk of the 
galaxy (Fig.~\ref{f:maps} top). As a consequence, in the present study we investigate star formation 
in a wider range of lengthscales, covering structures with sizes from few tens pc (clusters and small 
associations) to few hundreds pc (stellar aggregates and complexes). Apart from the study of hierarchy, 
our detection method allows us to also  address important issues of star formation, such as the 
structural behavior of the clusters in NGC~6822 and the {\sl cluster mass function} of the galaxy. 

The observational material, which includes  the most complete stellar sample observed in 
NGC~6822 \citep[][from here on dBW06]{deblok06} is described in \S~\ref{s:obsmater}. 
In \S~\ref{s:stelpop} we define the stellar content of the galaxy for the purpose of our study and select 
the type of stars our analysis focuses on. We identify the 
clusters of the galaxy with the application of the {\sl nearest neighbor} (from here on NN) method 
in \S~\ref{s:structident}, and we discuss their sizes and masses in \S~\ref{s:cluscat}. In \S~\ref{s:cmf}
we construct the {\sl cluster mass function} of NGC~6822, and compare our results with those of other
galaxies. The structural parameters of the detected clusters are measured and cross-correlated in
\S~\ref{s:structpar}. The hierarchical behavior in the spatial distribution of the blue stellar content
of NGC~6822 is investigated in \S~\ref{s:methstruct}, with the detection of stellar structures in
various length-scales larger than typical clusters, and the construction of their {\sl dendrogram},
to determine the degree of hierarchical clustering. We discuss our results concerning hierarchy in \S~\ref{s:concl}. 
Finally, we summarize the findings of this study in \S~\ref{s:sum}.



\section{Observational Material}\label{s:obsmater}

\subsection{Photometric Stellar Catalog}

In this investigation we use the photometric data derived from images of NGC~6822 taken with 
Suprime-Cam on the 8.2-m {\sc Subaru} Telescope at Mauna Kea, Hawaii. Suprime-Cam 
consists of 5~$\times$~2 CCDs of 2048~$\times$~4096 pixels each, providing a total 
field of view of 34\arcmin~$\times$~27\arcmin\ with a 0\farcs2 pixel size \citep{miyazaki02}. 
We make use of the results based on two sets of archival 
Suprime-Cam images observed in $B$, $R$ and $I$. The first set consists of two {\sl deep} 
pointings,  covering the whole of the {\sc Hi} disk of NGC~6822 
(see Fig.~\ref{f:maps} top). It was observed on 2001 October 15 and 19, and it is 
described in detail by \cite{komiyama03}. The second set of images was taken in 
2000 June, during the early days of Suprime-Cam, and it 
consists of a single {\sl shallow} pointing toward the optical center of NGC~6822. 
dBW06 stress the importance of the combination of the deep photometry with the shallow
for a complete understanding of the stellar population of NGC~6822, due to the high light 
gathering power of {\sc Subaru}, which saturates stars down to quite faint magnitudes.
The present study is based on the comprehensive photometric analysis of both data sets 
performed earlier in dBW06 for the investigation of the stellar content of NGC~6822. 
As a consequence, much of the information provided in this section comes from
the results of these authors. We repeat it here for reasons of completion.


\begin{figure*}[t!]
\centerline{\includegraphics[clip=true,width=0.55\textwidth]{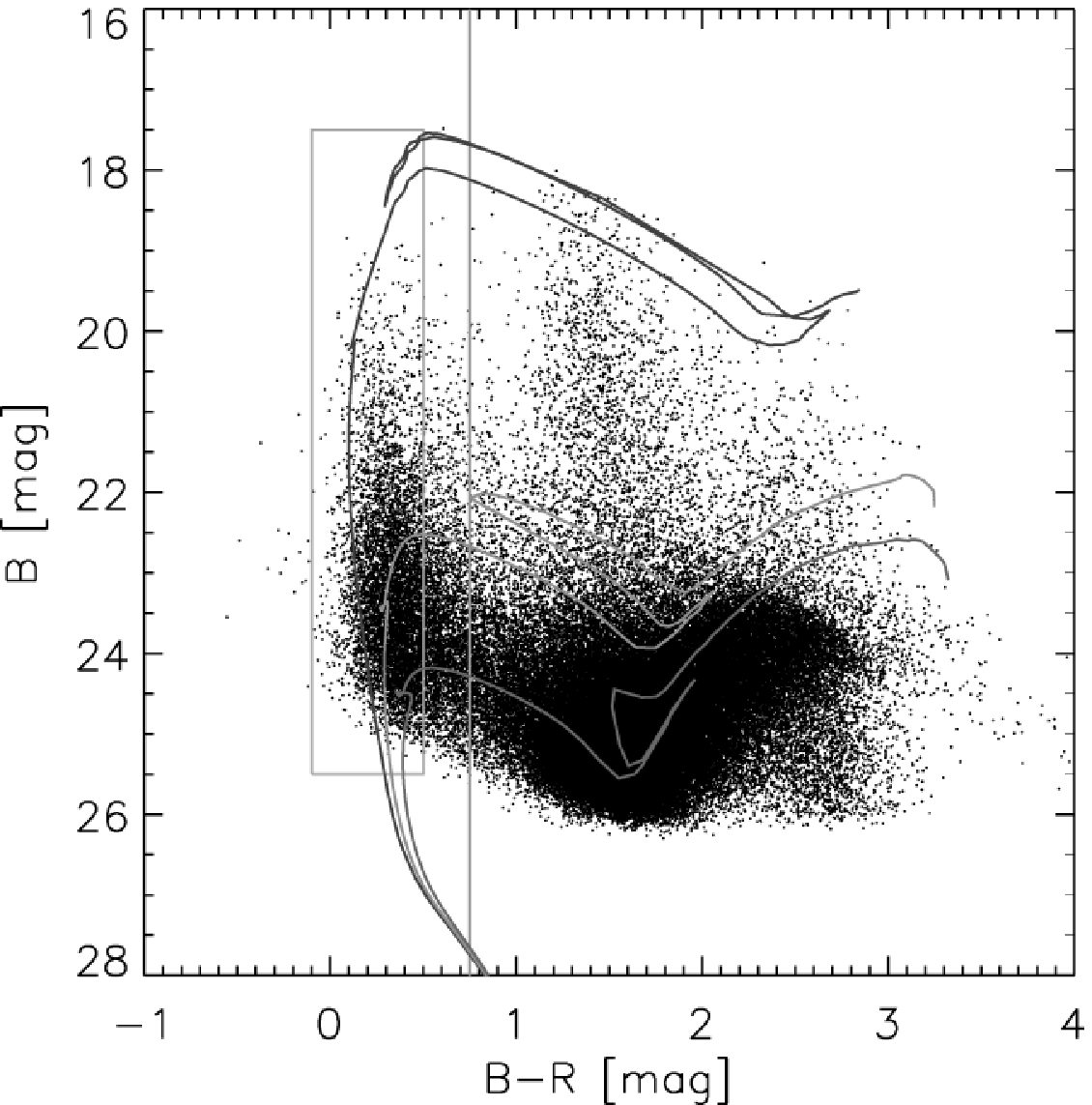}}
\caption{$B-R$, $R$ CMD of all 208~894 stars detected with 
good photometry in the observed {\sc Subaru} fields of NGC~6822.  
Indicative isochrones for 20 (blue), 200 (orange) and 500~Myr (red)
from the Padova evolutionary grid of models \citep{girardi02} are overlaid for a nominal
metallicity $Z=0.004$ \citep{skillman89}, reddening $E(B-V)=0.25$ \citep{massey07} and distance 
modulus $m-M=23.53$ \citep{gallart96a}. The green vertical line sets the limits of the ``blue plume''
stars, and the light blue box signifies the CMD area of the {\sl MS stars} with $(B-R) \leq 0.5$, 
selected in this study for the identification of young stellar structures with ages down to
roughly 500~Myr (see text in \S~\ref{s:stelpop}). [Color version of the figure will be available 
in the published electronic version of the paper.]
\label{f:cmd}}
\end{figure*}

\subsection{Color-Magnitude Diagram}

In the present study we utilize the final merged photometric catalog of stellar sources 
derived in dBW06 by combining the catalogs of stars detected in both deep 
and shallow exposures. This catalog contains 250~237 stellar sources identified 
in all three $B$, $R$, $I$ wavelengths.  In our analysis we limit ourselves to only those ``high-quality'' 
objects with a photometric uncertainty $\sigma \leq 0.1$~mag in all three bands, constraining 
the stellar catalog to 208~894 members. The $B$, $B-R$ color-magnitude diagram (CMD) of these stars 
is shown in Fig.~\ref{f:cmd}. This CMD has a larger dynamic range 
than those derived from previous investigations \citep{komiyama03, battinelli03, deblok03}, and 
that of SLGGCFS \citep{massey06, massey07}, 
probing to fainter magnitudes. By adding in the shallow data our stellar sample also covers the O- 
and B-type star regime of NGC~6822, and consequently it is the ideal stellar sample for the 
investigation of the most recent clustered star formation in this galaxy. 


\section{Young Stellar Populations in NGC~6822}\label{s:stelpop}

The CMD of Fig.~\ref{f:cmd} contains a mixture of different stellar populations. 
A complete discussion on all observed stellar types in NGC~6822 is given by dBW06.
In short this CMD shows three important distinct components: (i) The vertical 
``blue plume'' centered on $B-R \sim 0.3$, which consists mostly of 
young stars in NGC~6822. (ii) The red population of stars covering the area of
the CMD with 1~\lsim~$B-R$~\lsim~1.75 and $m_{B}$~\gsim~23 \citep[named the 
``red-tangle'' by][]{gallart96b}, which contains mainly the old and intermediate-age 
stellar content of NGC~6822 with ages between 1 and 10~Gyr.
(iii) A vertical band around $B - R \sim 1.5$, which consists of contaminant 
Galactic foreground stars \citep[see also][]{massey07}. Also clearly identified as 
being due to foreground stars is the region with $2 \lsim B-R \lsim 3$ and $m_{B} 
\gsim 24.5$ (dBW06). 

For our study we focus on the blue populations, and therefore 
we select our stellar sample from the ``blue plume''. This area in the CMD with 
$B-R < 0.75$ consists mostly of young main-sequence (MS) and more evolved 
helium-burning blue loop (BL) stars down to an age of roughly 0.5~Gyr. In the 
absence of variable foreground extinction as found toward NGC~6822, MS stars 
would occupy the blue side of the plume and BL stars the red. Therefore, since we 
are interested in the youngest observed populations, we constrain our selection 
of stars to those with $B-R \leq 0.5$, which correspond to ages roughly \lsim~500~Myr, 
according to the Padova evolutionary models \citep{girardi02}.

Fig.~\ref{f:maps} (middle) shows the spatial distribution of the selected sample of blue 
stars with $B-R \leq 0.5$. Since the final photometric catalog of dBW06 is the combined 
shallow photometric catalog in the inner part with the deep catalog in the outer 
part of the galaxy, these authors checked the outer field for saturated stars that should 
have been incorporated in an equivalent shallow catalog of the outer field. They found 
that there are very few saturated stars present, and therefore the stellar distribution 
shown in the middle panel of Fig.~\ref{f:maps} represents very well the whole observed  
blue stellar sample, despite the lack of a shallow photometry for the outer field. 
The corresponding surface density distribution of the blue MS stars constructed by simple 
star counts in a smoothed grid of elements 64\arcsec~$\times$~64\arcsec\ in size is shown in 
the bottom panel of Fig.~\ref{f:maps}. 

From the maps of Fig.~\ref{f:maps} one may conclude that the young blue
stars (i) are distributed across the whole {\sc Hi} disk, apart maybe from its 
southeastern part, (ii) they are mostly concentrated in the central part of the
galaxy in a ``S-like'' feature, and (iii) their distribution is rather clumpy. It should
be noted that the density contour map of Fig.~\ref{f:maps} (bottom) is a smoothed
visualization of the large-scale stellar structures  that exist in NGC~6822. The application,
later, of the NN method (\S~\ref{s:structident}) allows a far more detailed identification of
the smallest possible stellar concentrations of the galaxy. 


Interestingly, the star counts of Fig.~\ref{f:maps} revealed as an {\sl independent}
stellar structure the so-called ``Northwestern Cloud'', located at a position 
(R.A., Decl.)~$\approx (295.9, -14.6)$, which is speculated to be a separate 
system that is {\sl currently interacting} with the main body of NGC~6822  and is thought to be responsible 
for the tidal arms in the southeast of the galaxy \citep{deblok00, deblok03}. It could also have 
triggered the star formation that eventually led to the creation of the large hole in the southeastern 
part of the main {\sc Hi} disk, centered at a position (R.A., Decl.)~$\approx (296.35, -14.9)$ 
(dBW06).

The stellar content of the Cloud is dominated by a well-defined blue MS,
and there are two low-brightness {\sc Hii} regions identified in the Cloud (dBW06).
Both findings indicate that the Cloud has experienced recent star formation.
Unfortunately, the shallow {\sc Subaru} images do not cover its area, and therefore 
only stars up to $m_{B} \sim 20.5$ are measured, the brightest and bluest 
of which being well described by a $\log{\tau} \sim 7$ Padova isochrone.
The presence, however, of a few saturated stars clustered in the same way 
as the unsaturated ones and the H\alp\ observations suggest that even younger
stars may be present in the Northwestern Cloud (dBW06).

\begin{figure*}[t!]
\centerline{\includegraphics[clip=true,width=0.875\textwidth]{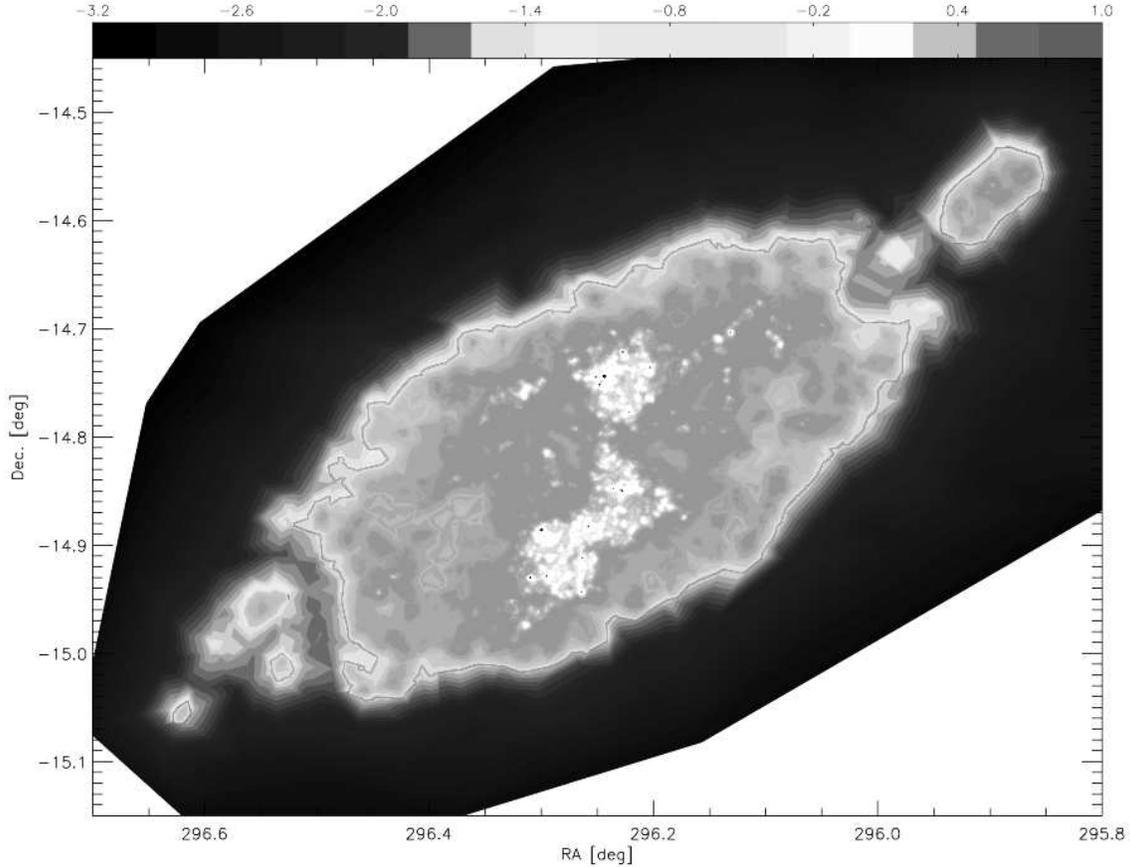}}
\caption{The stellar density map of NGC~6822 (in logarithmic scale), constructed with 
the NN density method applied to stars with ages \lsim~500~Myr for the 10th NN. All stellar 
concentrations detected with density higher than 3$\sigma$ above the average background 
are identified as members of the cluster population of NGC~6822. [Color version of the figure will be available 
in the published electronic version of the paper.]
\label{f:10nndenmap}}
\end{figure*}

\section{Detection of Concentrations of Young Stars}\label{s:structident}

The blue MS population selected in the previous section for the study of 
clustered star formation within the last $\sim$~500~Myr comprises 13~727 stars. 
In this section we apply the nearest neighbor (NN) method to investigate {\sl how 
these stars are clustered}. A short introduction of the method, which is  
described by, e.g., \cite{schmeja09}, is given below (\S~\ref{s:method}). 
Our method enables the detection of dense stellar concentrations, revealing the 
most prominent {\sl cluster population} of NGC~6822 (\S~\ref{s:methclus}). While these 
systems span an order of magnitude in physical dimensions, we refer to all of 
them as {\sl clusters}, although this term in principle applies only to a specific small-scale 
single-age class of systems, and not to the whole distribution of detected objects. 
Independent repetitions of the method, for different density thresholds 
(\S~\ref{s:methstruct}), revealed less dense stellar concentrations systematically 
belonging to larger ones, providing evidence of hierarchy in the distribution of the 
blue MS stars in NGC~6822. We refer to all these concentrations generally as 
{\sl structures}. 

\subsection{The Nearest Neighbor Density Method}\label{s:method}

Star clusters are usually identified as regions of a certain overdensity with respect to
 the background stellar density. The NN method, introduced by 
 \cite{casertano85} based on earlier work by \cite{vonhoerner63}, estimates the local 
 source density $\rho_j$ by measuring the distance from each object to its $j$th 
 nearest neighbor: \begin{equation} \rho_j = \frac{j - 1}{S(r_j)} m \end{equation}
where $r_j$ is the distance of a star to its $j$th nearest neighbor, 
$S (r_j)$ the surface area with the radius $r_j$ and $m$ the average
mass of the sources ($m = 1$ when considering number densities).
The chosen value of $j$ depends on the sample size, and is correlated 
with the sensitivity to the density fluctuations being mapped. A small $j$~value 
increases the locality of the density measurements while increasing the  
sensitivity to random density fluctuations. On the other hand a large $j$ 
value will reduce that sensitivity at the cost of some locality. 

The positions of the cluster centers are defined as the density-weighted enhancement centers 
\citep{casertano85}: \begin{equation} x_{{\rm d},j} = \frac{\sum_i x_i \rho_j^i}{\sum_i \rho_j^i},
\end{equation} where $x_i$ is the position vector of the $i$th cluster member and $\rho_j^i$
the $j$th NN density around this object. Similarly, the density radius $r_{\rm dens}$ is defined as the 
density-weighted average of the distance of each star from the density center: \begin{equation} 
r_{{\rm d},j} = \frac{\sum_i \vert x_i - x_{{\rm d},i} \vert \rho_j^i}{\sum_i \rho_j^i}. \label{eq:rad} \end{equation}
This radius typically corresponds to the core radius of the cluster \citep{casertano85}.

We applied the  NN method to our data considering several values for the nearest neighbors ($j$).
\cite{casertano85} have shown that low $j$ values, in particular $j = 1$ or 2, 
are extremely sensitive to statistical fluctuations, therefore they suggest using a value of $j \ge 6$.
On the other hand, the choice of a too large $j$ value results in a loss of sensitivity to real
density variations on smaller scales. 
 Monte Carlo simulations have shown that a value of $j = 20$ is adequate
 to the detection of clusters with about 10 to 1500 members (B. Ferreira, private communication).
However, since we aim at the detection of clusters even poorer than 10 members, we applied 
 several test runs of the method that showed that  $j = 10$ is a reasonable choice of 
 number of NN.

The 10th NN density map of the blue MS stars in NGC~6822 is shown in  
Fig.~\ref{f:10nndenmap}. This map demonstrates that the selected blue 
MS stars are distributed in almost the whole extent of the disk of the 
galaxy and the northern-western cloud, as it has been demonstrated by 
the iso-density contour map of Fig.~\ref{f:maps}. The NN density map, 
however, along with the general (average) stellar distribution of these stars 
throughout the galaxy, can also track in detail individual stellar concentrations. 
Specifically, while the low-density distribution of the blue MS stars, shown 
in blue and green scales, clearly outlines the disk of NGC~6822 and
the north-western cloud, the high-density levels (shown in yellow and red) 
show a more clumpy and centrally concentrated stellar distribution, revealing
individual stellar systems as density peaks.

\subsection{Detected Young Clusters}\label{s:methclus}

A reasonable density level, above which the detected density enhancements are accepted 
as star clusters is $3 \sigma$ above the average density, where $\sigma$ is the 
standard deviation of the background density. As a consequence, the original list of detected 
objects comprises all stellar density enhancements with density peaks at and above the 
3$\sigma$ threshold. However, since our method does not set any dimensional limit for the 
detected objects, this list includes the smallest density peaks that could be possibly identified.
Indeed, the smallest objects in the list are found to be minute density peaks, which include 
only one or two stars, and therefore they are most probably spurious detections at the level of 
the noise. Naturally, these detections cannot be related to real physical stellar concentrations, 
but rather to the  background density fluctuations. As a consequence, we exclude them from the 
final catalog of detected young clusters in NGC~6822, which thus comprises all stellar
concentrations of 3 or more stellar members, that are found with our method to have 10th 
NN density values $3 \sigma$ above the average density. 

The detected clusters 
contain MS stars in the whole observed magnitude range of $B$~\lsim~25. However, the range 
of brightness of the majority of the stars included in every cluster can be used to assign an indicative 
maximum age to each of them. We divided, thus, the observed catalog of MS stars into four magnitude 
ranges and we again computed the corresponding 10th NN density and applied a $3 \sigma$ threshold 
for the detection of clusters. Each of the selected magnitude ranges corresponds to a specific range of 
stellar ages, with the older age, specified by the {\sl turn-off} of the best-fitting 
isochrone, being associated to the faint magnitude limit. The four magnitude ranges were selected to 
include almost equal number of stars so that each run of the method would have the same statistical 
significance. We assign this age-limit to each of the stellar 
sub-catalogs with the use of the evolutionary models of \cite{girardi02} for the distance, metallicity 
and reddening of NGC~6822, by applying isochrone fitting of the CMD. In this manner we are able to 
have a direct estimate of the age-range of the clusters identified 
in the corresponding stellar sub-catalogs, with those detected in the whole catalog having the oldest
observable age of $\sim$~500~Myr. 

Although we cannot have an estimate of the actual age of each cluster, this approach is important for two 
reasons. First, age determination of the detected clusters by isochrone fitting of their individual CMDs is 
not really possible, due to their poor stellar number statistics. Moreover, the derivation of ages from 
integrated surface brightness of each cluster would require the application of population synthesis 
techniques, which would introduce important model-dependent uncertainties, again due to low stellar 
numbers, especially for the faint clusters. The 
selected magnitude ranges are $B <$~21, $B <$~22, $B <$~23 and $B \geq$~23, which according to 
the models correspond to maximum ages for the clusters of $\tau_{\rm max} \simeq$~95, 150, 250 and 
500~Myr respectively. The majority of the detected clusters (36) are found to comprise stars in the faintest
ranges or in the whole magnitude range and therefore  $\tau_{\rm max} \simeq$~500~Myr is assigned to 
them. Seven clusters are found with $\tau_{\rm max} \simeq$~250~Myr, one with $\tau_{\rm max} 
\simeq$~150~Myr, and three with $\tau_{\rm max} \simeq$~95~Myr. The maximum ages of the clusters 
are given in column 4 of Table~\ref{t:strpar} for comparison to their dynamical timescales derived in 
\S~\ref{s:structpar}. In the following Section we discuss in detail the characteristics of the detected clusters.

\section{The Cluster Population of NGC~6822}\label{s:cluscat}

The catalog of identified star clusters in NGC~6822 consists of  
47 clusters, the physical dimensions of which are defined by the 3$\sigma$ density level in the 
10th NN density map.  Table~\ref{t:cluslist} lists the identified clusters with the positions of 
their 10th NN density centers (Columns 2 and 3), the number of cluster members (Column 4), 
and their density and limiting radii (Columns 5 and 6 respectively) in seconds of an arc.  
The {\sl density radius}, $r_{\rm dens}$, (Column 5) is derived from the NN method
and Eq.~(\ref{eq:rad})  and roughly corresponds to the core radius of each system. The 
so-called {\sl effective radius} \citep[e.g.,][]{carpenter00} or {\sl equivalent radius} \citep[e.g.,][]{romanzuniga08}
of the systems (Columns 6 and 7) is the radius of a circle with the same area as the area enclosed by the 
cluster-defining 3$\sigma$ contour ($r_{\rm equiv} = \sqrt{A_{\rm cl}/\pi}$). The total {\sl expected} 
number of stars, $N$, (Column 10) and mass, $M_{\rm cl}$, (Column 11) are estimated from the extrapolation
of the mass spectrum of each cluster assuming that it reassembles the average Galactic IMF 
(see \S~\ref{s:clusmass}).

\begin{figure}[t!]
\centerline{\includegraphics[clip=true,width=0.5\textwidth]{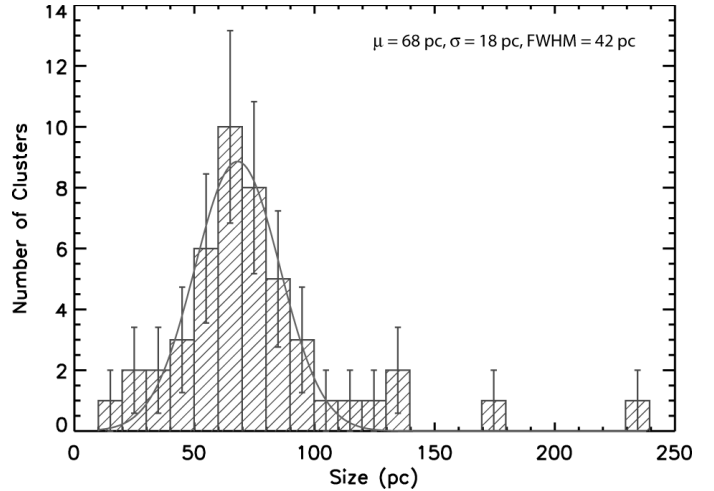}}
\caption{Histogram of the size distribution of detected young star clusters in NGC 6822, with a bin size of 
10 pc. Parameters of the best-fit Gaussian (red line) are also given. [Color version of the figure will be available 
in the published electronic version of the paper.]
 \label{f:sizedist}}
\end{figure}

\begin{deluxetable*}{rccrrrrrrrc}
\tablewidth{0pc}
\tabletypesize{\footnotesize}
\tablecaption{The Cluster Population of NGC~6822 with ages \lsim~500~Myr, 
detected with 10th NN density values $3 \sigma$ above background.\label{t:cluslist}} 
\tablehead{
\colhead{Cluster} & 
\colhead{R.A.} &
\colhead{Decl.} &
\colhead{$n_\star$} &
\colhead{$r_{\rm dens}$} &
\multicolumn{2}{c}{$r_{\rm equiv}$} &
\colhead{$B_{\rm br}$} &
\colhead{$m_{\rm max}$} &
\colhead{$N$}  &
\colhead{$M_{\rm cl}$}  
\\
\colhead{ID} & 
\colhead{(deg)} &
\colhead{(deg)} &
\colhead{} &
\colhead{(\arcsec)} &
\colhead{(\arcsec)} &
\colhead{(pc)} &
\colhead{(mag)} &
\colhead{(M{\solar})} &
\colhead{} &
\colhead{($10^3$~M{\solar})} 
} 
\startdata
1&	296.24692&	$-$14.74281	&	77&	14.70&	50.29&	119.47&	19.34	&	17.42	&	87812	&	$	9.99\pm	0.68$		\\
2&	296.22754&	$-$14.84399	&	47&	10.24&	37.68&	89.50&	18.94	&	19.33	&	62032	&	$	7.06\pm	0.49$		\\
3&	296.26373&	$-$14.91980	&	29&	7.60&	29.18&	69.32&	20.35	&	13.23	&	47708	&	$	5.43\pm	0.38$		\\
4&	296.25827&	$-$14.87879	&	24&	7.09&	29.02&	68.93&	19.37	&	17.27	&	48364	&	$	5.50\pm	0.39$		\\
5&	296.29990&	$-$14.88262	&	24&	4.42&	25.76&	61.20&	21.19	&	10.36	&	29269	&	$	3.33\pm	0.23$		\\
6&	296.29492&	$-$14.92559	&	19&	5.50&	24.38&	57.91&	22.38	&	7.17	&	34399	&	$	3.91\pm	0.28$		\\
7&	296.28711&	$-$14.89010	&	15&	7.31&	21.20&	50.37&	21.89	&	8.37	&	12538	&	$	1.43\pm	0.10$		\\
8&	296.26413&	$-$14.93961	&	15&	3.74&	20.20&	47.99&	21.52	&	9.37	&	34797	&	$	3.96\pm	0.29$		\\
9&	296.22186&	$-$14.77319	&	14&	4.04&	20.10&	47.76&	20.19	&	13.82	&	19757	&	$	2.25\pm	0.16$		\\
10&	296.20294&	$-$14.73244	&	16&	3.84&	19.99&	47.48&	19.30	&	17.58	&	16055	&	$	1.83\pm	0.13$		\\
11&	296.23712&	$-$14.72879	&	14&	4.39&	18.20&	43.23&	20.25	&	13.61	&	16998	&	$	1.93\pm	0.14$		\\
12&	296.29254&	$-$14.88720	&	11&	4.26&	17.64&	41.90&	21.92	&	8.29	&	18219	&	$	2.07\pm	0.15$		\\
13&	296.28781&	$-$14.93673	&	10&	6.29&	17.31&	41.12&	20.03	&	14.46	&	26106	&	$	2.97\pm	0.22$		\\
14&	296.23611&	$-$14.84454	&	10&	2.81&	17.24&	40.94&	21.49	&	9.46	&	21508	&	$	2.45\pm	0.18$		\\
15&	296.31003&	$-$14.92668	&	12&	2.13&	17.20&	40.86&	22.16	&	7.68	&	25761	&	$	2.93\pm	0.22$		\\
16&	296.24265&	$-$14.75290	&	10&	2.99&	16.30&	38.73&	20.82	&	11.55	&	21005	&	$	2.39\pm	0.17$		\\
17&	296.29214&	$-$14.90718	&	13&	5.14&	16.17&	38.40&	21.69	&	8.89	&	10096	&	$	1.15\pm	0.08$		\\
18&	296.22964&	$-$14.73439	&	10&	6.30&	16.01&	38.03&	20.95	&	11.14	&	10703	&	$	1.22\pm	0.08$		\\
19&	296.22729&	$-$14.71831	&	11&	1.67&	16.01&	38.03&	21.43	&	9.64	&	22769	&	$	2.59\pm	0.19$		\\
20&	296.27225&	$-$14.88502	&	10&	3.15&	15.72&	37.34&	21.09	&	10.67	&	34637	&	$	3.94\pm	0.30$		\\
21&	296.21454&	$-$14.76024	&	10&	4.29&	15.47&	36.75&	21.23	&	10.25	&	22983	&	$	2.61\pm	0.19$		\\
22&	296.23825&	$-$14.73670	&	9&	4.81&	15.46&	36.73&	20.75	&	11.80	&	12147	&	$	1.38\pm	0.10$		\\
23&	296.26706&	$-$14.92909	&	7&	3.51&	15.27&	36.28&	20.81	&	11.60	&	20109	&	$	2.29\pm	0.17$		\\
24&	296.26959&	$-$14.87903	&	9&	4.01&	14.69&	34.89&	22.70	&	6.45	&	14426	&	$	1.64\pm	0.12$		\\
25&	296.21567&	$-$14.84455	&	8&	4.55&	14.26&	33.88&	21.36	&	9.85	&	31207	&	$	3.55\pm	0.27$		\\
26&	296.13132&	$-$14.69955	&	11&	2.95&	14.06&	33.39&	19.58	&	16.35	&	38558	&	$	4.39\pm	0.35$		\\
27&	296.26364&	$-$14.90800	&	8&	2.39&	13.98&	33.22&	21.55	&	9.30	&	29133	&	$	3.31\pm	0.26$		\\
28&	296.27594&	$-$14.92400	&	7&	4.27&	13.55&	32.18&	21.59	&	9.18	&	9849	&	$	1.12\pm	0.08$		\\
29&	296.22586&	$-$14.77637	&	7&	3.80&	13.03&	30.95&	20.43	&	12.94	&	25421	&	$	2.89\pm	0.22$		\\
30&	296.27118&	$-$14.89139	&	7&	3.03&	12.82&	30.46&	23.03	&	5.79	&	17718	&	$	2.01\pm	0.15$		\\
31&	296.26489&	$-$14.87163	&	4&	1.76&	12.80&	30.42&	20.93	&	11.20	&	11498	&	$	1.31\pm	0.09$		\\
32&	296.22665&	$-$14.87727	&	9&	2.60&	12.79&	30.38&	21.38	&	9.79	&	26342	&	$	3.00\pm	0.23$		\\
33&	296.22006&	$-$14.84672	&	6&	2.38&	12.67&	30.09&	22.24	&	7.50	&	31063	&	$	3.53\pm	0.28$		\\
34&	296.22711&	$-$14.72315	&	8&	2.98&	12.33&	29.30&	20.86	&	11.41	&	26343	&	$	3.00\pm	0.23$		\\
35&	296.20230&	$-$14.76187	&	6&	4.90&	11.71&	27.81&	20.88	&	11.37	&	13942	&	$	1.59\pm	0.12$		\\
36&	296.19022&	$-$14.85646	&	7&	2.68&	11.49&	27.29&	19.84	&	15.22	&	18638	&	$	2.12\pm	0.16$		\\
37&	296.31396&	$-$14.97089	&	5&	2.49&	11.33&	26.90&	21.46	&	9.55	&	15824	&	$	1.80\pm	0.14$		\\
38&	296.20490&	$-$14.75966	&	5&	2.06&	11.11&	26.40&	21.70	&	8.88	&	9558	&	$	1.09\pm	0.08$		\\
39&	296.23492&	$-$14.72453	&	6&	2.86&	10.97&	26.06&	20.87	&	11.38	&	12427	&	$	1.41\pm	0.11$		\\
40&	296.23090&	$-$14.75460	&	6&	2.81&	10.37&	24.64&	20.92	&	11.22	&	17777	&	$	2.02\pm	0.16$		\\
41&	296.23541&	$-$14.73430	&	4&	2.33&	9.63&	22.87&	22.36	&	7.21	&	10157	&	$	1.15\pm	0.09$		\\
42&	296.19272&	$-$14.86216	&	3&	2.43&	9.00&	21.39&	22.82	&	6.22	&	12311	&	$	1.40\pm	0.11$		\\
43&	296.24420&	$-$14.73231	&	4&	1.62&	8.27&	19.65&	22.86	&	6.13	&	13818	&	$	1.57\pm	0.12$		\\
44&	296.31229&	$-$14.75217	&	4&	1.99&	7.41&	17.59&	21.29	&	10.05	&	11492	&	$	1.31\pm	0.10$		\\
45&	296.25198&	$-$14.88950	&	4&	2.42&	5.78&	13.74&	22.49	&	6.91	&	20460	&	$	2.32\pm	0.20$		\\
46&	296.24365&	$-$14.87206	&	3&	2.52&	5.58&	13.25&	21.42	&	9.66	&	14739	&	$	1.68\pm	0.14$		\\
47&	296.21973&	$-$14.88315	&	3&	4.02&	3.85&	9.14&	22.18	&	7.65	&	8670	&	$	0.99\pm	0.09$		
\enddata
\tablecomments{A detailed description of the parameters is given in \S~\ref{s:cluscat}.
\tablenotetext{}{}
}
\end{deluxetable*}

\subsection{Size Distribution of the Clusters}\label{s:clussize}

From Table~\ref{t:cluslist}  it can be seen that our catalog of young clusters covers a variety of 
systems, starting with those of the minimum of 3 stellar members with dimensions \lsim~20~pc
(cluster 47), up to those with maximum size of almost 240 pc, including 77 stars (cluster 1). 
The size distribution of the clusters constructed by binning them according to their dimensions
is shown in Fig.~\ref{f:sizedist}. Dimensions are given in physical units (pc) assuming a distance
modulus of $m-M = 23.53$. A functional fit to this histogram shows that the size distribution 
of the detected clusters can be very well approximated by a normal distribution. According to the 
best-fit Gaussian, drawn with a red line in Fig.~\ref{f:sizedist}, the dimensions of the detected 
systems are clustered around an average of $\simeq 68$~pc with a standard deviation of 
$\simeq 18$~pc. This is a rather interesting result, considering that, as pointed out earlier 
by various authors \citep[e.g.,][]{efremov87, ivanov96, gouliermis03}, young stellar 
associations {\sl in different galaxies} seem to have typical dimensions close to this value
(between 65 and 93 pc) with an average of 80~pc. If, however, this length-scale is characteristic
for star formation is still under debate \citep[see,e.g.,][]{bastian07, gouliermis09}.
 
 \subsection{Masses of the Clusters}\label{s:clusmass}

The sample of detected clusters naturally includes a variety of systems not only concerning their physical
dimensions but also their stellar content and in consequence their total mass. While our data are not deep
enough to cover the whole extent of masses for the stellar members of the detected clusters, and not complete
enough to provide us with the actual numbers of stars (per luminosity or mass range) for each of them, it is
worthwhile to perform a first-order calculation of the total stellar mass included in each detected cluster, based 
on several assumptions. We identify the number of stars included in 
each cluster and their luminosities from our photometric catalog. Then, in order to have a first estimation of 
the total mass of  each cluster we (i) build the observed Luminosity Function (LF) of each cluster, (ii) apply 
a mass-luminosity relation (M-LR) based on the stellar evolutionary models for the conversion of this LF to the 
Mass Function (MF) of the cluster, and (iii) extrapolate this observed MF to the lower masses for the calculation
of the total mass of each cluster.

\subsubsection{Mass-Luminosity Relation}

For the establishment of the conversion of stellar luminosities to masses, we use the evolutionary models by 
\cite{girardi02}. Our clusters identification is applied for all detected stars of NGC~6822 formed within the last 
$\sim$~500~Myr, and therefore the identified clusters are expected to have different ages from each other within
this time range. As a consequence, in order to construct a realistic M-LR for each cluster, one should establish 
a correct age with the use of the CMD of the detected stars in the cluster. However, the construction of individual M-LR 
for each cluster is not possible, because in all cases the stellar numbers found within the detected clusters are not
sufficient to build complete and informative CMDs. 

\begin{figure}[t]
\centerline{\includegraphics[clip=true,width=0.5\textwidth]{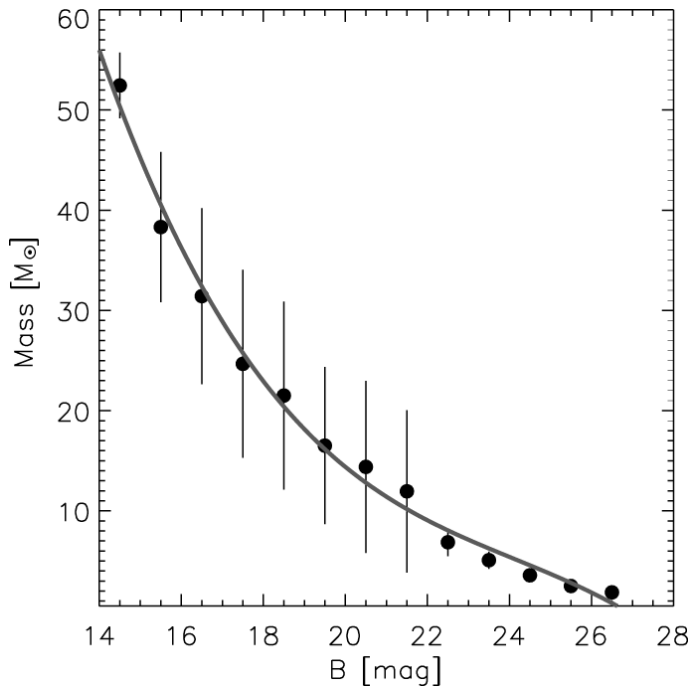}}
\caption{Mass-Luminosity relation constructed with the use of isochrones from the 
grid of evolutionary models by \cite{girardi02} for the metallicity and at the distance 
of NGC~6822.  Isochrones of a large variety in ages are used for the correct quantification of 
the uncertainties in the determination of the cluster masses due to age differences
among the identified clusters. [Color version of the figure will be available 
in the published electronic version of the paper.] \label{f:mlrat}}
\end{figure}

On the other hand, one may construct a {\sl global} M-LR to be used for the conversion of LFs to MFs 
for {\sl all} clusters, but this M-LR should be established from evolutionary models that cover a large 
range in ages. Indeed, as it is seen in the CMD of Fig.~\ref{f:cmd} the observed MS is populated by 
stars in an increasingly larger variety of ages towards fainter magnitudes down to $B\simeq25$~mag. 
Therefore, we construct a {\sl global}  M-LR for all identified clusters using the evolutionary 
models of \cite{girardi02} not for specific ages, but for all available ages and for the appropriate 
metallicity of $Z=0.004$.  With this treatment the 
uncertainties in stellar masses due to the age differences among the identified clusters will be 
inherited to the global M-LR and affect accordingly the determination of the total mass of the 
clusters. The derived M-LR, for an assumed average reddening $E(B-V)=0.25$ 
\citep{massey07} and distance modulus $m-M=23.53$ \citep{gallart96a}, is shown in Fig.~\ref{f:mlrat}.
From this figure it can be seen that indeed the 
stellar mass uncertainty is a function of the MS magnitude, with the larger uncertainties 
appearing for the parts of the MS where models of different ages show the greatest differences 
in the derived masses. While $B\simeq25$~mag corresponds more or less to our detection limit,
and our stellar sample does not include stars brighter than $B\simeq17.5$~mag, we construct 
our M-LR for a more extended magnitude range for reasons of completeness in its shape. This
M-LR is being used for the direct transformation of the luminosity of each star detected within 
the limits of one of the NGC~6822 clusters into mass.

\subsubsection{The Stellar Mass Spectrum of the Clusters}\label{s:sms}

The stellar mass function is the number distribution of stars according to their masses, 
constructed for a given volume of space in a cluster. This function is known as the stellar {\sl Present 
Day Mass Function}, which is usually called simply the {\sl Mass Function} of the 
cluster. In this study we refer to this function as the {\sl Stellar Mass Function} (SMF), to distinguish it
from the {\sl Cluster Mass Function} discussed later. 
The SMF of a cluster is described by the function $\xi(\log{m})$, which gives the number 
of stars per unit logarithmic (base ten) mass interval $d\log{m}$ per unit area (e.g. /kpc$^{2}$). 
Alternatively, one may refer to the {\sl Stellar Mass Spectrum}  (SMS) of the cluster, $f(m)$, 
which is the number of stars per unit mass interval $dm$ per unit area. The common use of 
both these functions is based on the definition of the stellar {\sl Initial Mass Function} (IMF), 
various forms of which are discussed by, e.g., \cite{kroupa02} and \cite{chabrier03}. All these distributions are
usually parametrized by their indices. Specifically, for the SMS $f(m)$, and the SMF 
$\xi(\log{m})$, these indices are defined as \beq \gamma = \frac{d\log{f(m)}}{d\log{m}} 
\eeq \label{eq-fga} and \beq \Gamma = \frac{d\log{\xi(\log{m})}}{d\log{m}}. 
\label{eq-xiga}\eeq These two indices  are basically the logarithmic derivatives or 
slopes of $f(m)$ and $\xi(\log{m})$ respectively and for power-law distributions they 
are independent of mass \citep{scalo86}. A reference value for the SMF slope is 
$\Gamma = -1.35$, which is the index of the classical IMF for stars in the solar 
neighborhood with masses 0.4 \lsim\ $M/M${\solar} \lsim\ 10, found by \cite{salpeter55}. The
corresponding SMS index is $\gamma = \Gamma - 1 \simeq -2.35$. For comparison, the
lognormal field-star IMF by \cite{miller79} has $\Gamma \simeq -(1+\log{m})$. A basic relation 
between SMF and SMS is $\xi(\log{m}) = (\ln{10})\cdot m \cdot f(m) \simeq 2.3 \cdot mf(m)$ 
\citep[see][]{scalo86}.

\begin{figure}[t]
\centerline{\includegraphics[clip=true,width=0.485\textwidth]{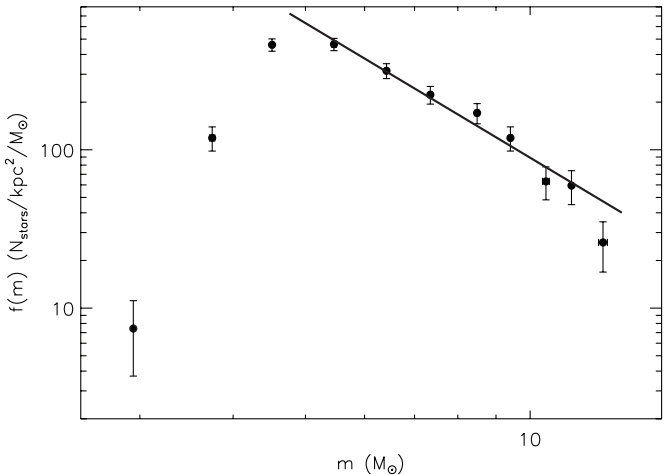}}
\caption{The Stellar Mass Spectrum of all stars comprised within the areas covered of all 47
detected clusters with age \lsim 500~Myr in NGC~6822. The total sample includes 561 stars
with masses between $\sim$~3 and 13~M{\solar}. The sample is complete for stars down to 
$\sim$~5~-~6~M{\solar}. A linear fit (solid line) to the SMS down to this mass limit 
gives a slope of $\gamma$ between  $-2.21$ and $-2.82$, very close to the typical Galactic field slope 
 for the same mass range. The plotted points correspond 
to the average mass of all stars counted per mass-bin with the x-axis errors derived from the standard
deviation of the masses per bin. The errors on the y-axis reflect the Poisson statistics. See 
also text in \S~\ref{s:sms}.\label{f:sms}}
\end{figure}

The construction of the SMS of each cluster is essential for the calculation of its {\sl expected} 
total mass through extrapolation of its SMS. However, in most of the cases of identified clusters
in NGC~6822 the detected stellar numbers are {\sl not} sufficient for the construction of a 
meaningful SMS for each cluster. This problem can be surpassed with the use of a common 
SMS constructed by all stars detected within the identified clusters. This approach, naturally, 
assumes that the SMS is indeed {\sl universal}, in line with most findings in the local universe
\citep[see, e.g., ][]{massey06a, bastian10}. We construct the {\sl global} SMS of the clusters in NGC~6822 
by counting all stars detected within the limits of all the 47 identified clusters according to their 
masses, and distributing the stellar masses in mass-bins 1~M{\solar} wide. This SMS  
is given in Table~\ref{t:sms}, and it is shown in Fig.~\ref{f:sms}. This figure also shows the best-fit 
line derived from the application of a linear regression to the most complete mass-bins corresponding 
to masses between $\sim$~5 and 13~M{\solar}.

\begin{deluxetable}{rrrcc}
\tablewidth{0pc}
\tablecaption{The global {\sl Stellar Mass Spectrum}  constructed from all stars detected within the limits
of the 47 identified clusters in NGC~6822.\label{t:sms}}
\tablehead{
\colhead{$\langle m_{\rm bin} \rangle$} &
\colhead{$N_{\rm stars}$} &
\colhead{$\langle m \rangle$} &
\colhead{$\sigma_{\langle m \rangle}$} &
\colhead{$\rho_{\star}$} \\
\colhead{M{\solar}}&
\colhead{}&
\colhead{M{\solar}}&
\colhead{M{\solar}}&
\colhead{stars/kpc$^2$/M{\solar}}
}
\startdata
2.50    &       1       &       2.94    &       -    &       $7.43    \pm       3.71$    \\
3.50    &       31      &       3.75    &       0.03    &       $118.83  \pm       20.68$   \\
4.50    &       123     &       4.51    &       0.02    &       $460.48  \pm      41.19$   \\
5.50    &       124     &       5.46    &       0.04    &       $464.19  \pm       41.35$   \\
6.50    &       84      &       6.42    &       0.01    &       $315.65  \pm       34.04$   \\
7.50    &       59      &       7.35    &       0.04    &       $222.81  \pm       28.52$   \\
8.50    &       45      &       8.49    &       0.05    &       $170.82  \pm       24.91$   \\
9.50    &       31      &       9.41    &       0.04    &       $118.83  \pm      20.68$  \\
10.50   &       16      &       10.50   &       0.11    &       $63.13   \pm       14.85$   \\
11.50   &       15      &       11.36   &       0.04    &       $59.42   \pm       14.38$   \\
12.50   &       6       &       12.52   &       0.17    &       $25.99   \pm       9.10$   
\enddata\
\tablecomments{The mean mass per bin, given in Col. 1, is comparable to the measured average 
mass of Col. 3, calculated from all stars counted per bin, given in Col. 2. The corresponding 
standard deviations, given in Col. 4, do not exceed the width of each bin (1~M{\solar}). The 
counted stellar numbers per surface used for the construction of the SMS, and the corresponding 
Poisson errors are given in Col. 5. Values are plotted in Fig.~\ref{f:sms}.
\tablenotetext{}{}
}
\end{deluxetable}

The determination of the slope of power laws from uniformly binned data using linear regression
is found to comprise biases, which are caused by the correlation between the number of stars per 
bin and the assigned weights \citep{maizapellaniz05}. However, we measured the average mass 
of the stars counted in each bin for the construction of the global SMS, and we found that it is 
almost equal to the mean mass of the bin. The corresponding standard deviations per bin, are 
found to be smaller than 1~M{\solar}, the width of each bin, and therefore there is no ``leak'' of 
stars expected from one mass-bin to its neighboring, and thus no random fluctuations in the 
constructed SMS. The {\sl measured} average mass per bin, rather than the mean mass
of each bin, is plotted in Fig.~\ref{f:sms}, and the corresponding $\sigma_{\langle m \rangle}$ 
are shown as error bars on the x-axis, demonstrating the accuracy of the pointing of each mass-bin.
The linear regression for the most complete mass range of 5~\lsim~$m$/M{\solar}~\lsim~13 
provides a SMS slope $\gamma$, which is exceptionally close to the average Galactic field 
value, equal to $-2.5 \pm 0.5$ (black-solid line in Fig.~\ref{f:sms}).

\subsubsection{Total Masses of the Clusters}\label{s:totmas}

\begin{figure}[t!]
\centerline{\includegraphics[clip=true,width=0.485\textwidth]{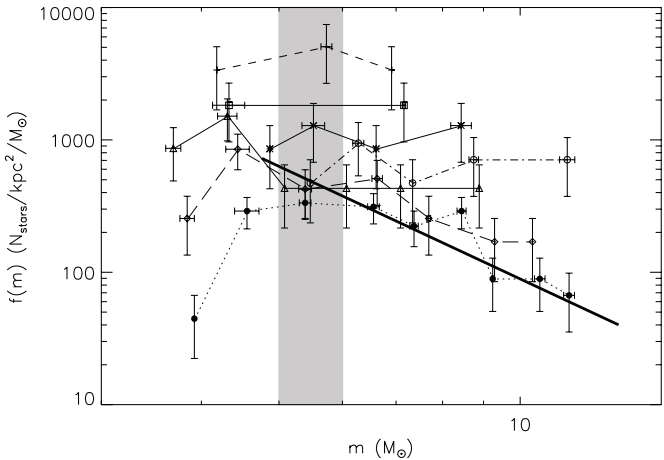}}
\caption{Indicative Stellar Mass Spectra (SMS) from the sample of clusters detected 
in NGC~6822. Different symbols are used for plotting the SMS of seven indicative clusters, 
comprising a variety of objects from the richest cluster in the sample (Cluster \#~1) plotted with 
filled circles to one of the poorest (Cluster \# 41) with only two mass-bins plotted with open boxes. 
Points for individual clusters are also connected together with lines. The
slope of the global SMS is represented by the solid line, as in Fig.~\ref{f:sms}. This graph demonstrates
the variability in stellar numbers among the clusters, and the inability to extrapolate the observed SMS 
of each cluster, due to low stellar numbers and thus the necessity for the use of a global SMS. The mass 
range over which the global SMS is considered for extrapolation for all clusters, $m \simeq$~5~-~6~M{\solar}, 
is indicated by the vertical shaded area. See also text in \S~\ref{s:totmas}
\label{f:sms-smpl}}
\end{figure}

An estimate of the total mass of each cluster in NGC~6822 can be achieved through integration of
its SMS. However, as we discuss earlier, the construction of a meaningful  SMS for each cluster
is not possible due to low stellar numbers, and therefore we make use of the global SMS assuming 
that this distribution is universal and representative of all clusters. While, according to our assumption, 
the slope of the SMS should not be different from one cluster to the other, the absolute stellar numbers
per mass-bin of the SMS in each cluster are different and thus the corresponding total cluster mass. 
Indeed, the individual SMS of the clusters do deviate from the global SMS, mostly in stellar numbers
and, in some cases of clusters with very few stars, in the slope. This is demonstrated in Fig.~\ref{f:sms-smpl},
where the SMS of seven selected indicative clusters from our sample are plotted. In this figure it can be seen
that the general trend of the SMS slope indeed seems consistent with the global SMS (solid line), but
the stellar numbers (per surface unit) are quite different. These differences will determine the differences in
total mass of the clusters. We estimate the cluster masses in three steps: 

\begin{itemize}

\item[(i)] We first extend the observed global SMS to the low-mass regime below our detection limit 
according to the generally accepted Galactic field mass spectrum. This mass spectrum is usually 
parametrized with a series of power laws, with exponents changing in different 
mass ranges \citep[see, e.g., ][]{scalo98}. The most recent parameterization is that by \cite{kroupa02},
according to which the slope $\gamma$ changes from $\gamma = -0.3$ in the substellar mass range, to $\gamma 
= -1.3$ for masses between 0.08~M{\solar} and 0.5~M{\solar}, $\gamma = -2.3$ for $0.5 \leq m/{\rm M}{\solar} < 1$, 
and $\gamma$ between $-2.7$ and $-2.3$ ($\pm 0.3$) for stars of higher masses. This SMS is generally characterized 
as the Galactic average, in the sense that it is reasonably valid for different regions of the Galaxy.

Considering that the SMS of all clusters identified in NGC~6822 should have a multi-power law form, 
with exponents that change at the same mass limits as the Galactic SMS, we assume that the global 
SMS of the clusters has the average slope of those measured in \S~\ref{s:sms}, equal to $\gamma = -2.5 \pm 0.5$,
for stars with masses down to our completeness limit of about 5.5~M{\solar} and we extrapolate it with 
the same slope down to the mass limit of 1~M{\solar}. For the extrapolation of the 
global SMS to sub-solar masses we consider the exponents of the Galactic SMS, as they are discussed
above for stars with masses down to $\sim$~0.1~M{\solar}.

The extrapolated global SMS, following the parameterization by \cite{kroupa02}, is thus a three-part power law
function of the form \beq f(m) \propto~{\left(\frac{m}{m_i}\right)}^{\gamma_i}~{\rm with}~i=1,2,3. \label{eq:globsms} \eeq 
The slopes $\gamma$ depending the mass range are 
\beq 
          \begin{array}{l@{\quad,\quad}ll@{\quad}}
\gamma_1 = -1.3\pm0.5   &0.1 &\le m/{\rm M}_\odot < 0.5 \\
\gamma_2 = -2.3\pm0.3   &0.5  &\le m/{\rm M}_\odot < 1.0\\
\gamma_3 = -2.5\pm0.5  &1.0 &\le m/{\rm M}_\odot. \\
          \end{array}
\label{eq:smsslopes}
\eeq


\item[(ii)] We then normalize the extrapolated global SMS to fit the stellar numbers observed in every cluster. 
Specifically, we normalize the extrapolated global SMS to the mass-bin of around 5 to 6~M{\solar}, because 
all detected clusters, including those with only three stars, show to have at least one star at this specific mass 
range. Therefore, by normalizing the global SMS to this mass limit we apply a comparable approach to all, 
rich and poor, clusters. 

\item[(iii)] We finally integrate the normalized SMS of each cluster from the largest mass observed 
in the cluster to the smaller stellar mass assumed to exist in all clusters of 0.1~M{\solar}, and derive 
an estimation of the total cluster mass, from its surface stellar density 
\beq M_{\rm cl} =   \rho_{\star}~\pi~r_{\rm equiv}^2, \label{eq:totmas} \eeq 
where 
\beq \rho_{\star} = \int_{0.1}^{m_{\rm max}} f(m)~dm.\label{eq:intsms} \eeq 

\end{itemize}
The total stellar numbers of the clusters derived from this method and the corresponding
total stellar masses are given in Table~\ref{t:cluslist} (Cols. 10 and 11 respectively). 

\section{Mass Distribution of the Clusters}\label{s:cmf}

\begin{figure}[t!]
\centerline{\includegraphics[clip=true,width=0.485\textwidth]{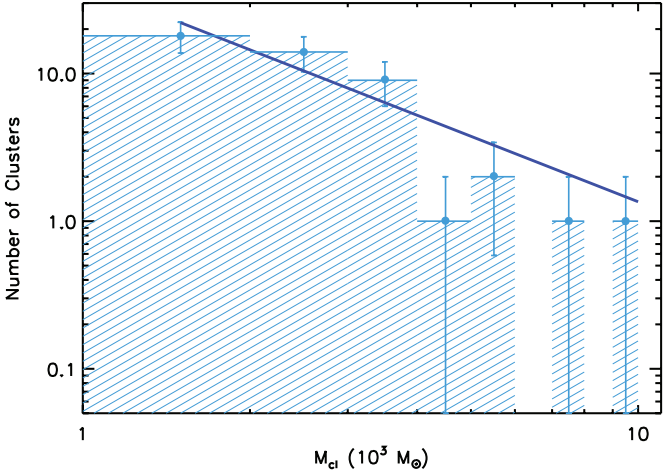}}
\caption{The Mass Spectrum of the star clusters or the {\sl Cluster Mass Function} 
of NGC~6822, based on calculation of their total stellar masses in \S~\ref{s:totmas} 
under the assumption that all clusters follow a common high-mass stellar IMF, 
extrapolated to the sub-stellar limit according to the {\sl Average Galactic} IMF.
[Color version of the figure will be available 
in the published electronic version of the paper.]
\label{f:cmf}}
\end{figure}

The total stellar masses of the clusters of NGC~6822, as derived above, cover a range
over an order of a magnitude with the smaller cluster mass being around $10^3$~M{\solar}.
The number distribution of the clusters according to their masses, i.e. the mass spectrum 
of the clusters, which is widely termed the {\sl Cluster Mass Function} (CMF) is shown in 
logarithmic scale in Fig.~\ref{f:cmf}. The CMF derived from young ({\lsim}10~Myr)
star clusters is defined as  the {\sl Cluster Initial Mass Function} (CIMF). Considering 
that the CMF follows the functional form of a power-law, as it is generally assumed, we 
applied a linear regression to the data of Fig.~\ref{f:cmf}. The best fit returns the function 
\beq  \displaystyle \log{N_{\rm cl}} = (-1.47 \pm 0.72)~\log{M_{\rm cl}} + (1.60 \pm 0.31). \eeq 
We found, thus, that the CMF of the massive clusters of NGC~6822 can be described 
by a power law with index $\sim -1.5 \pm 0.7$.

\subsection{Comparison with the CIMF}

Previous works on the CIMF in a variety of galactic environments have shown that this 
function is well described by a power law with index $-2$ \citep[see, e.g.,][]{gieles09}:
\beq \frac{d N}{d M} \propto M^{-\alpha}~,~~~\alpha = 2~. \eeq Specifically,
the CMF of star clusters in the Antennae galaxies (NGC~4038/9) with ages \lsim~160~Myr
is found, over the range $10^4 \lsim M_{\rm cl}/{\rm M}{\solar} \lsim 10^6$, to have an index of $-2$ 
\citep{zhang99}. \cite{mccrady07}  constructed the CMF of 19 super star clusters (SSCs) with 
$M_{\rm cl} > 10^5$~M{\solar} and ages of the order of 10~Myr in the nuclear starburst of M~82, 
and found a power-law with an index $-1.91 \pm 0.06$. The SSCs with an average age $\sim$~10~Myr 
in the nearby starburst galaxies NGC~3310 and NGC~6745 are investigated by \cite{degrijs03}, who find 
CMF indexes of $-2.04\pm 0.23$ and $-1.96\pm 0.19$ respectively. These CMFs correspond to the CIMF 
due to the youthfulness of the cluster samples. All three studies on massive star clusters agree to an almost 
identical CIMF index of $-2$. In addition, a similar CIMF index of $-2.1\pm 0.3$ is found for 10~Myr old 
clusters in the mass range $2.5~10^{3} < M_{\rm cl}/{\rm M}{\solar} < 5~10^{4}$  in the inner spiral arms 
of M~51 \citep{bik03}. All the aforementioned studies provide evidence of a ``universality'' of the CIMF. 
Since most of the clusters in our sample have undergone substantial dynamical and stellar evolution, 
their CMF cannot be compared with the CIMF. It is interesting, though, to notice that the derived CMF 
index of NGC~6822 in the mass range $10^3$~-~$10^4$~M{\solar} of $\sim$~$-1.5\pm0.7$ 
is well within 1$\sigma$ of the ``global'' CIMF, as the large uncertainties in our CMF do not allow
us to identify any significant difference from it.


\subsection{Comparison with the Magellanic Clouds}

Our CMF of NGC~6822, which covers stellar concentrations with ages of up to $\tau\simeq 500$~Myr, 
should have been significantly affected by evolutionary effects and therefore it is not directly comparable 
to the CIMF. Comparable CMFs in terms of cluster masses and ages to that derived here are 
those presented by \cite{degrijs06} and \cite{degrijs08} for the cluster populations in the Large 
and Small Magellanic Clouds (LMC, SMC) respectively. The CMF index in the LMC is given 
as a function of minimum mass and maximum age, based on mass-limited samples \citep[see][their 
Table 3 and Fig.~8]{degrijs06}. For $\log{(m_{\rm cl}/{\rm M}{\solar})}_{\rm min} = 3$ and 
$\log{(\tau/{\rm yr})}_{\rm max} = 8.7$, which corresponds to our sample, \cite{degrijs06} find a
CMF index $-1.98\pm0.08$, an average value which is somewhat different than ours, but nevertheless 
covered by our uncertainties, and quite similar to the CIMF. In addition, in the SMC \cite{degrijs08} find
a CMF  with index $\alpha$~\lsim~$1.2$ for almost the same cluster mass range with ours. This slope falls 
within 1$\sigma$ of our CMF slope, and is therefore also consistent with it. 

\section{Structural Parameters of the Clusters}\label{s:structpar}

We estimate the stellar density, $\varrho_{\rm cl}$, in the half-mass radius,
$r_{\rm h}$, of each cluster and the disruption time, $t_{\rm d}$, of the cluster 
due to interaction with passing-by interstellar clouds, as described in, e.g., \citet[][their
section 6]{gouliermis02}. The former is calculated assuming spherical symmetry 
for all clusters as \beq \displaystyle \varrho_{\rm cl} \equiv \frac{M_{\rm cl}}{\frac{4}{3}\pi 
r_{\rm h}^{3}}~~\bigg( \frac{{\rm M}_{\odot}}{{\rm pc}^{3}}\bigg),\eeq while the latter is 
given as \citep[][]{spitzer58} \begin{equation} t_{\rm d} = 1.9 \times 10^{8} 
\varrho_{\rm cl}~~({\rm years}). \label{eq:td} \end{equation} 

The dynamical status of a young star cluster is defined by two additional 
time-scales, the {\sl crossing} and the {\sl two-body relaxation} time, which are given as 
\citep[e.g.,][]{kroupa08} \begin{equation} t_{\rm cr} \equiv \frac{2 r_{\rm
h}}{\sigma}~~~{\rm and}~~~t_{\rm relax} = 0.1 \frac{N}{\ln{N}} t_{\rm
cr} \label{eq:tcr}\end{equation} respectively. The three-dimensional
velocity dispersion of the stars in the cluster, $\sigma$, is given as
\begin{equation} \sigma = \displaystyle{ \sqrt{ \frac{{\rm G}M_{\rm
cl}}{\epsilon r_{\rm h}}},} \end{equation} where $\epsilon$ is the
star formation efficiency (SFE) and $r_{\rm h}$ the half-mass radius of
the cluster. We estimate these time-scales for the clusters of our sample,
with the application of Eqs.~(\ref{eq:tcr}), assuming the same SFE with 
several nearby Galactic gas-embedded clusters, which has been found 
to range typically from 10\% to 30\% \citep{lada03}. In our calculations we 
assume an average value of $\epsilon \simeq 0.2$, which leads to values 
of $\sigma$ between 12 and 89~km~s$^{-1}$ for our clusters. 

For the measurement of the $r_{\rm h}$ of each cluster we estimate the so-called {\sl Spitzer} radius, 
$r_{\rm Sp}$. Considering that $r_{\rm h} \simeq 0.9~r_{\rm Sp}$ only for 
dynamically relaxed spherical clusters \citep[e.g.,][]{spitzer69}, we can only 
approximate the actual $r_{\rm h}$ of our clusters. The Spitzer radius is a 
dynamically stable radius, defined by the mean-square distance of the 
stars from the center of the cluster \citep[e.g.,][]{spitzer87} as \begin{equation} 
\displaystyle{r_{\rm Sp} \equiv \sqrt{\frac{\displaystyle{\Sigma_{i=1}^{N}} 
r_{i}^{2}}{N} }}, \label{eq:reff}\end{equation} where $r_{i}$ is the projected 
radial distance of the $i$th star of the cluster in a total sample of 
$N$ stars. We estimate this radius for each cluster in our sample 
only from the stars identified within its limits, for which there are positions
available, and not from the expected total number of stars according to the 
extrapolated mass spectrum of each cluster. As a consequence, the accuracy 
of the derived $r_{\rm Sp}$ of each cluster is subject to the number 
of its identified stars.

The estimation of $r_{\rm Sp}$ allows the evaluation of $\sigma$ and $\varrho_{\rm cl}$
of the clusters and consequently of their $t_{\rm d}$, $t_{\rm cr}$ and $t_{\rm relax}$ 
according to equations~(\ref{eq:td}) and (\ref{eq:tcr}). For the latter we use the {\sl expected} total number
of stellar members, $N$, as it is derived from the extrapolation of the mass 
spectrum of each cluster. All additional estimated structural 
parameters for the young clusters of NGC~6822 are 
provided also in Table~\ref{t:strpar}.

\begin{deluxetable*}{rccrrrcrrr}
\tabletypesize{\footnotesize}
\tablewidth{0pc}
\tablecaption{Structural parameters of the star clusters of NGC~6822.\label{t:strpar}} 
\tablehead{
\colhead{Cluster} & 
\colhead{R.A.} &
\colhead{Decl.} &
\colhead{$\tau_{\rm max}$}  &
\colhead{$\varrho_{\rm cl}$} &
\colhead{$r_{\rm Sp}$}  &
\colhead{$\sigma$}  &
\colhead{$t_{\rm d}$}  &
\colhead{$t_{\rm cr}$}  &
\colhead{$t_{\rm relax}$}  
\\
\colhead{ID} & 
\colhead{(deg)} &
\colhead{(deg)} &
\colhead{(Myr)}  &
\colhead{(M{\solar}/pc$^3$)} &
\colhead{(pc)}  &
\colhead{(km~s$^{-1}$)}  &
\colhead{(Myr)}  &
\colhead{(Myr)}  &
\colhead{(Myr)}  
} 
\startdata
1&	296.24692&	$-$14.74281	&	500&$	0.05	\pm	0.00	$	&	36.92	&	89.0	&	9.00	&	0.81	&	626.83		\\
2&	296.22754&	$-$14.84399	&	500&$	0.09	\pm	0.01	$	&	26.87	&	63.8	&	16.49	&	0.82	&	463.66		\\
3&	296.26373&	$-$14.91980	&	500&$	0.14	\pm	0.01	$	&	21.16	&	49.7	&	25.97	&	0.83	&	369.59		\\
4&	296.25827&	$-$14.87879	&	500&$	0.11	\pm	0.01	$	&	22.79	&	51.9	&	21.07	&	0.86	&	385.82		\\
5&	296.29990&	$-$14.88262	&	500&$	0.33	\pm	0.02	$	&	13.48	&	31.1	&	61.66	&	0.85	&	242.05		\\
6&	296.29492&	$-$14.92559	&	500&$	0.21	\pm	0.02	$	&	16.40	&	37.1	&	40.14	&	0.87	&	285.19		\\
7&	296.28711&	$-$14.89010	&	500&$	0.05	\pm	0.00	$	&	18.43	&	23.8	&	10.32	&	1.52	&	201.97		\\
8&	296.26413&	$-$14.93961	&	500&$	0.63	\pm	0.05	$	&	11.42	&	31.2	&	120.41	&	0.72	&	238.97		\\
9&	296.22186&	$-$14.77319	&	500&$	0.39	\pm	0.03	$	&	11.11	&	23.2	&	74.29	&	0.94	&	187.72		\\
10&	296.20294&	$-$14.73244	&	150&$	0.30	\pm	0.02	$	&	11.28	&	21.0	&	57.75	&	1.05	&	174.13		\\
11&	296.23712&	$-$14.72879	&	250&$	0.25	\pm	0.02	$	&	12.31	&	22.6	&	47.05	&	1.07	&	186.06		\\
12&	296.29254&	$-$14.88720	&	500&$	0.32	\pm	0.02	$	&	11.54	&	22.7	&	61.08	&	1.00	&	185.32		\\
13&	296.28781&	$-$14.93673	&	500&$	0.18	\pm	0.01	$	&	15.92	&	31.9	&	33.31	&	0.98	&	251.41		\\
14&	296.23611&	$-$14.84454	&	250&$	1.25	\pm	0.09	$	&	7.76	&	20.2	&	237.77	&	0.75	&	162.28		\\
15&	296.31003&	$-$14.92668	&	500&$	2.16	\pm	0.16	$	&	6.87	&	20.8	&	410.02	&	0.65	&	164.18		\\
16&	296.24265&	$-$14.75290	&	250&$	0.97	\pm	0.07	$	&	8.37	&	20.7	&	184.95	&	0.79	&	166.92		\\
17&	296.29214&	$-$14.90718	&	500&$	0.07	\pm	0.01	$	&	15.45	&	19.5	&	14.13	&	1.55	&	169.78		\\
18&	296.22964&	$-$14.73439	&	95&$	0.06	\pm	0.00	$	&	16.53	&	20.8	&	12.23	&	1.56	&	179.65		\\
19&	296.22729&	$-$14.71831	&	500&$	3.48	\pm	0.26	$	&	5.62	&	17.7	&	660.40	&	0.62	&	141.35		\\
20&	296.27225&	$-$14.88502	&	500&$	1.44	\pm	0.11	$	&	8.67	&	27.1	&	273.73	&	0.63	&	207.82		\\
21&	296.21454&	$-$14.76024	&	500&$	0.36	\pm	0.03	$	&	11.97	&	25.9	&	69.20	&	0.90	&	206.96		\\
22&	296.23825&	$-$14.73670	&	500&$	0.14	\pm	0.01	$	&	13.46	&	20.0	&	25.71	&	1.32	&	170.37		\\
23&	296.26706&	$-$14.92909	&	95&$	0.71	\pm	0.05	$	&	9.15	&	21.2	&	135.34	&	0.85	&	171.56		\\
24&	296.26959&	$-$14.87903	&	500&$	0.30	\pm	0.02	$	&	10.94	&	19.6	&	56.74	&	1.09	&	164.55		\\
25&	296.21567&	$-$14.84455	&	500&$	0.41	\pm	0.03	$	&	12.76	&	31.2	&	77.53	&	0.80	&	241.65		\\
26&	296.13132&	$-$14.69955	&	500&$	2.19	\pm	0.17	$	&	7.82	&	27.2	&	416.40	&	0.56	&	206.02		\\
27&	296.26364&	$-$14.90800	&	500&$	2.36	\pm	0.18	$	&	6.94	&	22.2	&	449.03	&	0.61	&	173.41		\\
28&	296.27594&	$-$14.92400	&	500&$	0.23	\pm	0.02	$	&	10.50	&	15.9	&	43.85	&	1.29	&	138.64		\\
29&	296.22586&	$-$14.77637	&	500&$	0.72	\pm	0.06	$	&	9.88	&	24.8	&	136.12	&	0.78	&	195.75		\\
30&	296.27118&	$-$14.89139	&	500&$	0.93	\pm	0.07	$	&	8.03	&	18.6	&	176.13	&	0.84	&	153.02		\\
31&	296.26489&	$-$14.87163	&	500&$	2.88	\pm	0.21	$	&	4.77	&	11.6	&	548.05	&	0.81	&	99.22		\\
32&	296.22665&	$-$14.87727	&	250&$	2.23	\pm	0.17	$	&	6.84	&	21.0	&	424.44	&	0.64	&	165.29		\\
33&	296.22006&	$-$14.84672	&	500&$	3.21	\pm	0.25	$	&	6.40	&	22.0	&	610.64	&	0.57	&	170.91		\\
34&	296.22711&	$-$14.72315	&	500&$	1.56	\pm	0.12	$	&	7.72	&	22.3	&	295.83	&	0.68	&	175.53		\\
35&	296.20230&	$-$14.76187	&	250&$	0.20	\pm	0.02	$	&	12.45	&	20.6	&	37.25	&	1.18	&	173.04		\\
36&	296.19022&	$-$14.85646	&	500&$	1.49	\pm	0.12	$	&	6.97	&	17.8	&	283.35	&	0.77	&	145.36		\\
37&	296.31396&	$-$14.97089	&	500&$	1.55	\pm	0.12	$	&	6.52	&	15.9	&	294.40	&	0.80	&	131.66		\\
38&	296.20490&	$-$14.75966	&	250&$	1.37	\pm	0.10	$	&	5.74	&	11.6	&	260.40	&	0.97	&	101.32		\\
39&	296.23492&	$-$14.72453	&	95&$	0.80	\pm	0.06	$	&	7.49	&	15.1	&	152.41	&	0.97	&	128.27		\\
40&	296.23090&	$-$14.75460	&	500&$	1.10	\pm	0.09	$	&	7.59	&	18.2	&	209.69	&	0.82	&	148.76		\\
41&	296.23541&	$-$14.73430	&	500&$	1.16	\pm	0.09	$	&	6.20	&	12.4	&	219.38	&	0.98	&	107.88		\\
42&	296.19272&	$-$14.86216	&	500&$	1.44	\pm	0.11	$	&	6.14	&	13.6	&	273.66	&	0.89	&	115.82		\\
43&	296.24420&	$-$14.73231	&	500&$	3.20	\pm	0.25	$	&	4.89	&	12.8	&	607.40	&	0.75	&	108.19		\\
44&	296.31229&	$-$14.75217	&	500&$	1.31	\pm	0.10	$	&	6.20	&	13.2	&	248.96	&	0.92	&	113.13		\\
45&	296.25198&	$-$14.88950	&	500&$	1.93	\pm	0.17	$	&	6.60	&	18.2	&	366.74	&	0.71	&	146.81		\\
46&	296.24365&	$-$14.87206	&	500&$	1.52	\pm	0.13	$	&	6.41	&	15.2	&	289.23	&	0.83	&	126.88		\\
47&	296.21973&	$-$14.88315	&	250&$	0.25	\pm	0.02	$	&	9.78	&	14.4	&	47.81	&	1.33	&	127.30			
\enddata
\tablecomments{The assignment of an indicative maximum age (Col. 4) for each cluster is discussed in \S~\ref{s:methclus}. The 
estimation of the structural parameters of the clusters is described in detail in \S~\ref{s:structpar}.
\tablenotetext{}{}
}
\end{deluxetable*}

\subsection{Parameters correlations of the clusters in NGC~6822}\label{s:parcorel}

\begin{figure}[t!]
\centerline{\includegraphics[clip=true,width=0.485\textwidth]{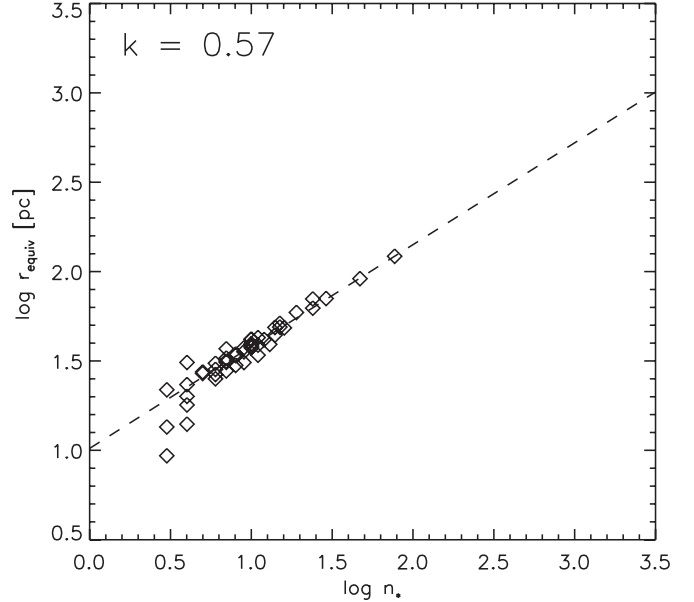}}
\caption{Correlation between the number of detected stars and $r_{\rm equiv}$ for the cluster population of NGC~6822 
(10th NN, 3$\sigma$ density detection level). This correlation is consistent with that found for Galactic embedded 
clusters (see \S~\ref{s:parcorel}). \label{f:n_vs_radius}}
\end{figure}

A correlation between the number of stars in a cluster and the radius of the cluster
has been found in Galactic embedded clusters \citep{adams06, allen07}:
While $n_*$ and $r$ vary by about two orders of magnitude, the average surface
density of cluster members ($n_*/\pi r^{2}$) is nearly constant.
We find the same correlation, following a power law, between $n_*$ and $r_{\rm equiv}$
for the clusters identified in NGC~6822, which span an even wider range in radius,
$n_*$, and age (Fig.~\ref{f:n_vs_radius}). The correlation has a slope in the 
log-log plot of $k \simeq 0.6$, comparable to that found by \cite{adams06} for their sample 
of young Galactic clusters. Our fit, which has a correlation coefficient of $q \approx 0.92$, 
is affected by the smallest clusters ($n_* \le 5$) that show some scatter. 
A similar correlation exists also between $n_*$ and $r_{\rm dens}$, but it 
becomes less clear when using the 
expected number of stars $N$ instead of the observed number $n_*$. A correlation
between number of stars and radius is also observed for the stellar structures we 
identify below with the application of the NN method with density levels different than 3$\sigma$ 
(see \S~\ref{s:strucorel}).

\section{Hierarchical clustering in NGC~6822}\label{s:methstruct}


The majority of stars form in clusters and aggregates of various sizes and masses \citep{lada03, allen07}. 
Young star clusters are often found inside larger complexes, which can be parts of even larger structures.
In turn, stellar clusters often consist of distinct subclusters, appearing to form a hierarchy of 
systems over a wide range of scales \citep[e.g.][]{efremov+elmegreen98, elmegreen00}. The ISM is also 
hierarchically structured (sometimes described as fractal) in scales starting from the largest giant molecular 
cloud down to individual clumps and cores. Within this scheme, the cluster population of 
NGC~6822, as presented above, should represent one specific length-scale of a whole spectrum of 
stellar concentrations in this galaxy, and the detected clusters themselves  are  
most probably members of larger structures. In this section we uncover the complete structural spectrum 
of stellar clustering and quantify the hierarchical distribution of the blue stars in NGC~6822 with 
$\tau$~\lsim~500~Myr, as they are selected in \S~\ref{s:stelpop}.


\begin{figure*}[t!]
\centerline{\includegraphics[clip=true,width=0.775\textwidth]{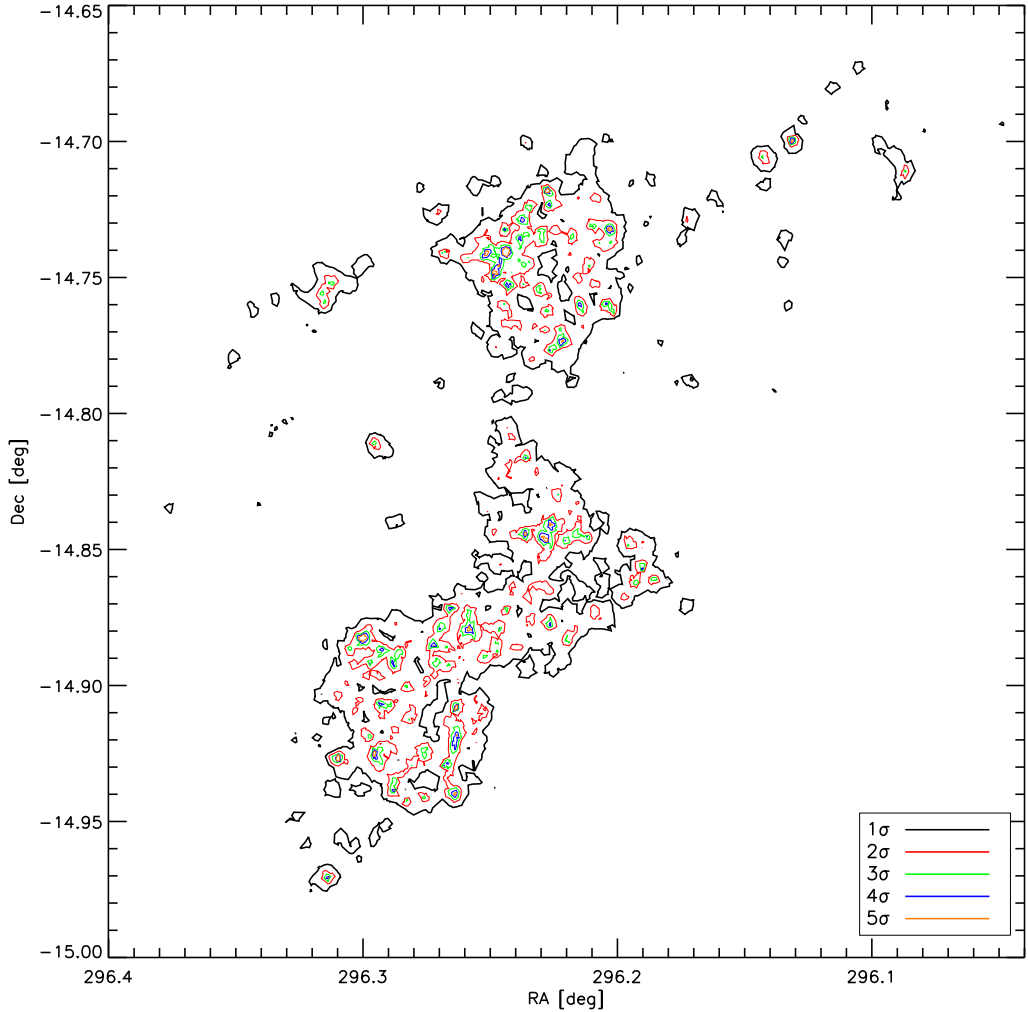}}
\caption{The density contour map constructed with the application of the 10th NN density method on our photometric 
data of NGC~6822. Isopleths within different  density levels in $\sigma$, drawn with different colors, signify 
the corresponding identified stellar structures in the galaxy. All statistically important stellar structures 
are detected only in the main part of NGC~6822 disk. Therefore, the drawn contour map covers only this 
part of the galaxy instead of the whole observed field-of-view for reasons of clarity. \label{f:contmap}}
\end{figure*}

\subsection{Detection of stellar structures in NGC~6822}

In order to search for stellar concentrations of various length-scales 
we apply again the NN density method, as described in \S~\ref{s:method}, for the 10th nearest neighbor 
(NN). The selection of the 10th NN is made for reasons of consistency with the detection of star clusters 
in the previous sections. Our tests showed that the 
application of the NN method for larger number of neighbors (e.g., 30 or 50) would `smooth' the
derived structures and unify many of them into single larger ones, loosing the ability to detect any fine-structure
in the spatial distribution of stars.
For the identification of stellar structures of different densities we select different density thresholds in the
application of the NN method. Specifically, apart from the 3$\sigma$ detection applied in \S~\ref{s:cluscat} 
for the identification of the cluster population of the galaxy, we apply the method for the lower density thresholds 
of 1$\sigma$ and 2$\sigma$ above the average background density level to identify structures that correspond to 
lower stellar density enhancements. 
For the detection of concentrations that correspond to high stellar density we apply the 
NN method  with higher (4$\sigma$ and 5$\sigma$) density threshold above the background level. 

The stellar structures identified with the 10th NN density method are shown in the density contour map
of Fig.~\ref{f:contmap}. Isopleths of different colors indicate the structures derived with the application of 
the method for different detection density thresholds. The cluster population of the 3$\sigma$ detection
is shown with green contours. This map is constrained to the main part of the galaxy, where all significant 
stellar concentrations, including its clusters, appear and which is known to host star formation \citep{cannon06}. 
From the map of Fig.~\ref{f:contmap} it can be seen that the low density threshold detections reveal  large stellar 
structures, while the application of the method for higher density thresholds give rise to smaller and more compact 
stellar concentrations, which are actually located {\sl in} the larger structures of the galaxy. This combination 
of high-density enhancements that correspond to small stellar systems and aggregates with low-density 
concentrations that represent large structures and stellar complexes is a clear signspot  of hierarchical structure 
\citep{elmegreen10}.

\subsection{Dendrograms}\label{s:dendro}

An intuitive way to illustrate hierarchical structures is through the so-called {\sl dendrograms}, introduced as ``structure trees'' 
for the analysis of molecular cloud structure by \cite{houlahan92}, refined by \cite{rosolowsky08}.
A dendrogram is constructed by cutting the image at different thresholds and identifying connected 
areas, while keeping track of the connection to ``parent structures'' (on a lower level) and ``child structures'' 
(on the next higher level, lying within the boundaries of its parent). A geometrically perfect hierarchy would be 
represented by a dendrogram where each parent branches out into the same number of children at each level.

\begin{figure}[t!]
\centerline{\includegraphics[clip=true,width=0.485\textwidth]{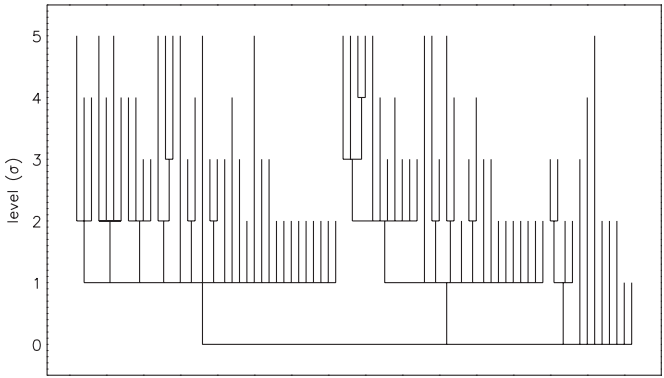}}
\caption{Dendrogram of the stellar structures identified at different levels in the 10th NN density 
map (Fig.~\ref{f:contmap}), illustrating the hierarchical behavior of the clustering of the blue MS stars.
\label{f:dendrogram}}
\end{figure}

We construct the dendrogram of the stellar structures detected in NGC~6822 with the NN density method
for the various density thresholds considered, i.e., 1 to 5$\sigma$ above the background density. In this dendrogram, 
shown in Fig.~\ref{f:dendrogram}, the structures found at each density level  are represented not only by the `leaves' 
that end at the particular level, but by all branches present  at that level. For example, at the 3$\sigma$ level, there 
are 47 branches of the  dendrogram (regardless whether they end here, continue to a higher level, or split into two 
or more branches), corresponding to the 47  detected stellar clusters. This dendrogram demonstrates that, while 
there are few centrally concentrated 
structures with a single peak, most structures split up into several substructures over at least three levels.
The combination of this dendrogram with the contour map of Fig.~\ref{f:contmap} illustrates graphically 
the hierarchical spatial distribution of stars younger than 500~Myr in NGC~6822.

In the lower detection density threshold of 1$\sigma$ (black isopleths in Fig.~\ref{f:contmap}) there are 
few structures revealed, three of them being major stellar concentrations that qualify as large {\sl stellar 
complexes}. These
structures comprise a number of smaller {\sl multiple} concentrations seen in the 2$\sigma$ red 
isopleths of Fig.~\ref{f:contmap}, which most probably correspond to the so-called {\sl stellar aggregates}
of the galaxy. The detected clusters (green 3$\sigma$ isopleths) are actually members of these aggregates,
fulfilling the typical image of hierarchical structuring of stars in a galaxy-scale. Higher density (4 and 5$\sigma$) 
detections correspond to the condensed stellar density peaks, which are seen to be the most compact centers of the 
larger structures.

\subsection{Quantification of hierarchy}

Another popular method in studies of hierarchical clustering is the {\sl Minimum Spanning 
Tree} (MST) method, which may be applied with different cutoff lengths for the detection of 
different length-scales  \citep[e.g.,][]{bastian07, bastian09}. The MST method is excellent in 
quantifying the hierarchy in the discovered structures through the so-called \Q\ parameter 
\citep[see e.g.,][]{cw04, sk06}. It allows to distinguish between clusters 
with a central density concentration and hierarchical stellar concentrations with possible 
fractal substructure. Large \Q\ values ($\Q > 0.8$) describe centrally condensed clusters 
having a volume density $n(r) \propto r^{- \alpha}$, while small \Q\ values ($\Q < 0.8$)
indicate concentrations with fractal substructure. \Q\ is correlated with the radial density 
exponent $\alpha$ for $\Q > 0.8$ and anti-correlated with the fractal dimension $D$ for 
$\Q < 0.8$. We determine \Q\ for all structures detected  at all levels with more than 45 
members, since for poorer concentrations the method becomes unreliable due to 
increasing errors. Recent simulations of artificial clusters have shown that for clusters with $N 
\leq 45$~members, the expected error in the determination of the \Q\ parameter 
is of the order of less than 0.05 \citep{schmeja10}. \Q\ varies systematically for structures showing a more elongated 
shape \citep{cw09, bastian09}, therefore we apply the correction suggested by 
\cite{bastian09} when necessary.

\begin{figure*}[t!]
\centerline{\includegraphics[clip=true,width=0.75\textwidth]{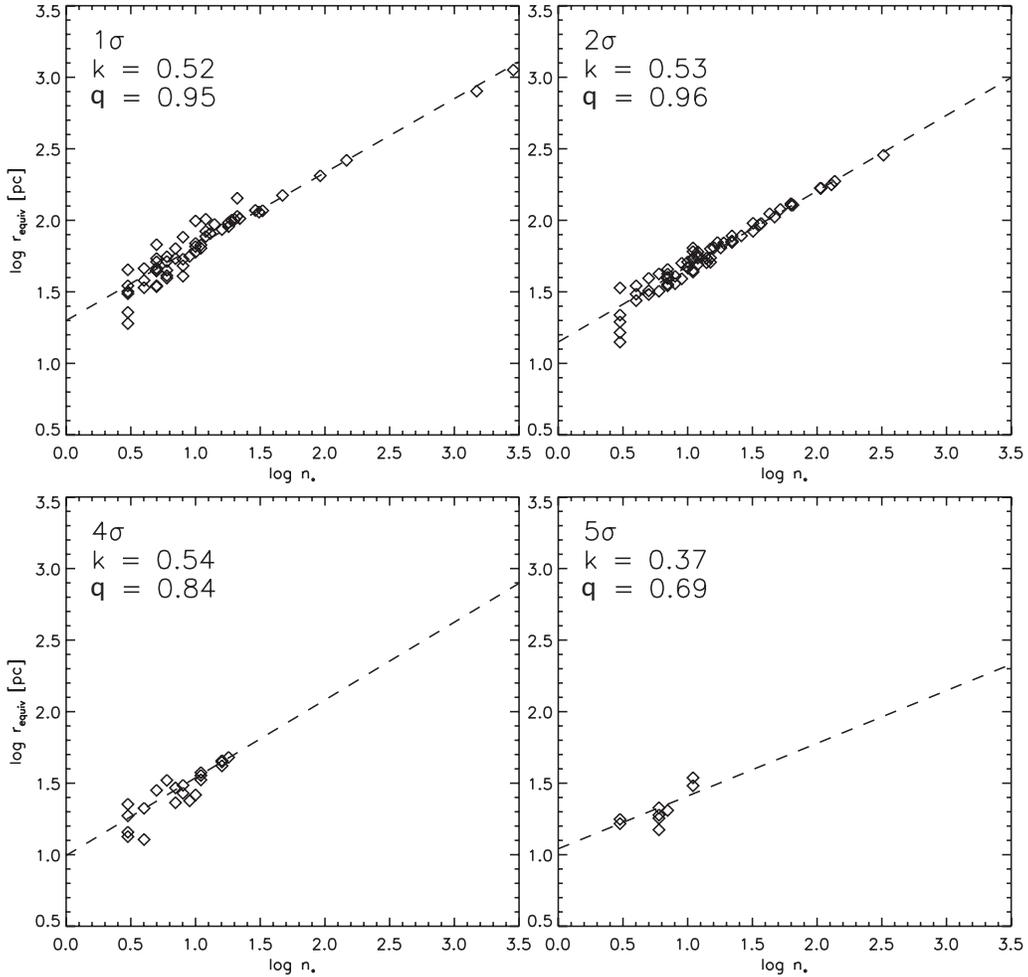}}
\caption{Correlation between the number of detected stars and $r_{\rm equiv}$ for the stellar structures of 
NGC~6822 identified with the NN method for various density detection levels. The corresponding slopes, $k$,
and correlation coefficients, $q$, are also given. This correlation is similar
to that we find for cluster population of the galaxy (\S~\ref{s:parcorel}) and consistent with findings for 
Galactic clusters (see \S~\ref{s:parcorel}). \label{f:nvsradius}}
\end{figure*}

The \Q\ values for the structures detected at 1, 2, and $3 \sigma$ density thresholds  
are all but one in the hierarchical regime ($\Q < 0.8$), reaching \Q\ values as low
as $\Q = 0.55$, corresponding to a fractal dimension of $D \approx 1.8$. The detections 
at higher density thresholds did not provide stellar structures rich enough to allow
such an analysis. The \Q\ values for the five 1$\sigma$ structures with more than 45 
stellar members range from 0.56 to 0.73, and those for the ten 2$\sigma$ structures from 
0.55 to 0.83. There are only two clusters detected in the 3$\sigma$ density threshold 
that include more than 45 stars, clusters No. 1 and 2 in Table~\ref{t:cluslist}. They are
also found to be hierarchically structured with values of $\Q = 0.62$ and 0.77 respectively.
The average \Q\ value for the structures revealed in all three density levels 
give evidence of fractal concentrations with $\bar{\Q} = 0.70$.
This value demonstrates the hierarchical nature of the clustering in NGC~6822. 
All but one of the analyzed structures are classified as hierarchical, i.e., they have 
significant (possible fractal) substructure rather than a smooth density gradient.
It should be noted that the numbers of stars in the objects revealed at 
density levels with thresholds $\geq$~3$\sigma$ are too low to permit a 
meaningful \Q\ analysis, but taking as example the two star clusters, 
for which this analysis was possible, most -- if not all -- of the 3$\sigma$ clusters 
should be expected in the hierarchical regime as well.

Nevertheless, the unusual distribution of blue stars in the central part of 
NGC~6822 and their apparent hierarchy may be the result of differential 
extinction. For example, the two vacant areas to the east and west of the 
central S-shaped distribution of blue stars (Fig.~\ref{f:contmap}), may be caused by high 
levels of extinction. Indeed, a comparison of the spatial distribution of the 
cool component of neutral hydrogen with that of the blue stars shows an 
anticorrelation on large scales (dBW06, their Fig. 7). However, while this 
extinction seems to affect the large-scale stellar distribution (at 1$\sigma$ 
density level), giving it the S-shape, it does not seem to affect the 
hierarchical distribution of smaller stellar structures detected in higher 
density levels (2 and 3$\sigma$). Concerning the smallest detected structures 
(at 4 and 5$\sigma$), the resolution of 
the available observations of the ISM in NGC 6822 does not allow the detection 
of any patchy extinction in smaller scales. As such, we cannot assess how 
small-scale differential extinction may affect the apparent hierarchy of 
stellar structures, and therefore we certainly cannot exclude it as a possible 
bias.

In this case the derived \Q\ values would indicate that the stars were in fact 
not in total hierarchically spread. It is also worth noting that if the largest 
stellar structures, shown in Fig.~\ref{f:contmap}, are approximately 2D (i.e., much longer 
and wider than deep) the \Q\ values returned by the MST method should be 
interpreted differently, since the crossover between `fractal' and `centrally 
concentrated' in the 2D case becomes 0.72 \citep[see,][]{bastian09}, and thus 
the structures may be found to be less hierarchical.

\subsection{Correlation of number of stars and radii}\label{s:strucorel}

The correlation between the number of stars and radius seen for the clusters 
of NGC~6822 detected at the 3$\sigma$ level (\S~\ref{s:parcorel}) is also
observed for the structures we identify with the application of the NN method
with different detection levels. This correlation is shown in Fig.~\ref{f:nvsradius}
for detections with density thresholds of 1, 2, 4 and 5$\sigma$. The number of stars
correlates with the radii of the structures with about the same slope of 
$k \approx$~0.53. Only the $5 \sigma$ structures show a different
behavior, but this plot suffers from the small number of objects.
\cite{adams06}  found a slope of $k = 0.54$ for  young Galactic clusters, 
very close to the slope we find for the detected stellar structures and clusters in 
NGC~6822. As pointed out by \cite{allen07}, these results are similar to 
that of Larson's relations for molecular clouds, $\rho \sim R^{-1}$ \citep {larson81}, 
implying a constant column density of the gas in molecular clouds.

\subsection{Current Star formation in NGC~6822}

\cite{cannon06} performed {\sl Spitzer} Space Telescope imaging of NGC~6822 and 
combined it with H$\alpha$, {\sc Hi} and radio continuum observations to study the 
nature of the emission in these wavebands on spatial scales of $\sim$~130~pc.
They found strong variations in the relative ratios of H$\alpha$ and IR flux 
throughout the central disk of the galaxy, and that the localized ratios of dust to 
{\sc Hi} gas are about a factor of 5 higher than the global value of NGC~6822.
These authors identified 16 IR emission complexes in the central part of the galaxy,
six of which correspond to detections in the radio continuum image. The major {\sc Hii} 
regions, i.e., the strongest H$\alpha$ sources of the galaxy are also luminous in the IR 
and identified  among the IR emission complexes, but in general H$\alpha$ and IR 
emission are not always cospatial. The {\sc Hi} distribution in the same area contains 
mostly high surface brightness (column densities \gsim~$10^{21}~{\rm cm}^{-2}$), and
it is clumpy with dense comps of neutral gas surrounding various emission peaks in other 
wavebands.

\begin{figure}[t!]
\centerline{\includegraphics[clip=true,width=0.485\textwidth]{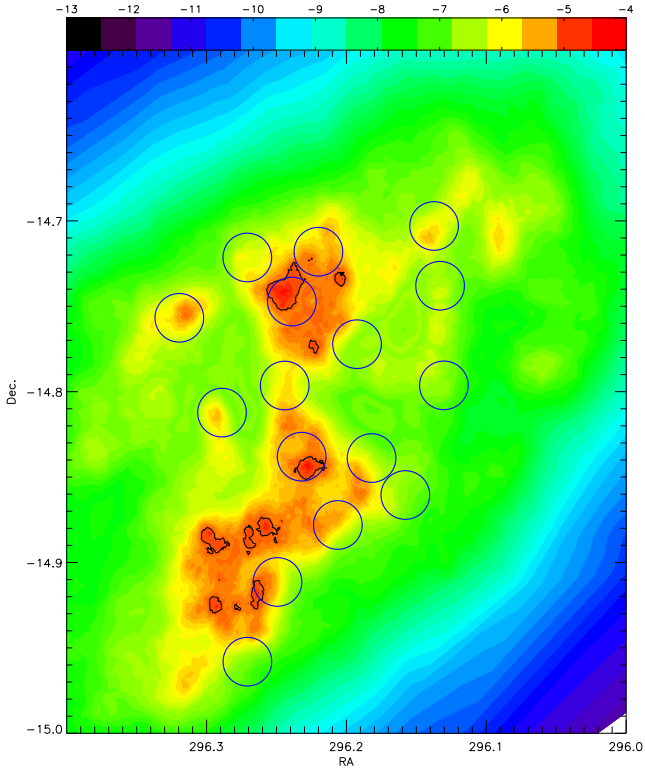}}
\caption{The 50th NN density map of the blue MS stars in the central region of NGC~6822 (in logarithmic scale)
with blue circles overlaid indicating the IR-bright complexes identified by \cite{cannon06} with {\sl Spitzer}.
\label{f:spitzer_comp}}
\end{figure}

It is worthwhile to compare the star-forming complexes identified by \cite{cannon06} 
with the structures revealed from our study. The IR emission complexes are indicated by
circles in Fig.~\ref{f:spitzer_comp}, overlaid on our NN density map of the central part of 
NGC~6822 disk. This map appears more smooth than those shown before because we constructed 
it by applying the NN density method for the more gross number of the 50th NN, in order to 
achieve an angular resolution comparable to that provided by the nebular and dust emission 
observations. The comparison of our stellar structures with the IR-bright regions of Cannon et al. 
on the map of Fig.~\ref{f:spitzer_comp} shows a correlation of the stellar density
peaks with the maximum IR emission only in the northern part of the galaxy, 
in particular, their regions 9 and 10 coincide with peaks in the stellar density.
In the southern part they are significantly displaced from each other.  Here, 
current star formation takes place at the western rim of the region of
high stellar density. This is reminiscent of the spatial distribution of stellar
birth in spiral arms or tidal tails, and points towards an external interaction  
\citep[see also][]{karampelas09}. On the other hand, this spatial disagreement 
between peaks in IR emission and in stellar density in the southern central region
of NGC~6822 may simply suggest that young stars are not yet revealed 
in their natal clusters, and we only detect the most evolved stellar
concentrations in this part of the galaxy.

\section{Discussion and conclusive remarks}\label{s:concl}

The formation of stars and star clusters in the interstellar matter (ISM) is
controlled by the complex dynamical interplay between self-gravity and supersonic
turbulence \citep{maclowklessen04}. On large scales, the turbulent motions in the
ISM are highly supersonic with Mach numbers up to several tens in giant molecular
clouds. On scales of individual star-forming cloud cores, the turbulent velocity
field becomes subsonic, with a well-defined power-law connecting these scales
\citep{larson81} which could be the result of the scale-free nature of the turbulent
cascade \citep{elmegreenscalo04, scaloelmegreen04}. The resulting
morphological structure has often been interpreted in terms of hierarchical fractals
(\citealp[e.g.,][]{elmegreenfalgarone96, stutzki98}; see also \citealp{federrath09, federrath10}). 

Turbulence plays a dual role. On global scales it provides support,
while at the same time it can promote collapse locally. It creates strong density
fluctuations with gravity taking over in the densest and most massive regions, where
collapse sets in to build up individual stars \citep{klessen01a}. Together with the
thermodynamic properties of the gas and magnetic fields this regulates the
fragmentation behavior in star forming clouds. The properties of young star clusters
are thus directly related to the statistical characteristics of the underlying
turbulence in the star-forming gas \citep{ballesteros-paredes07}. This holds for
the stellar IMF \citep{klessenburkert00, klessenburkert01, hennebelle08, 
hennebelle09}, the protostellar accretion rates 
\citep{klessen01b, schmeja04}, the star-formation efficiency and timescale 
\citep{klessenetal00, heitsch01, vazquez-semadeni03, krumholz05}, or the 
question of when and where dense clusters with high-mass stars form
and when and where to expect a more distributed population of lower-mass stars 
\citep[e.g.,][]{vazquez-semadeni09}.

Quite a number of different measures  have been proposed to support and
statistically characterize the link between the morphological and kinematical
structure of the ISM and the population of star clusters that form in it. 
\cite{efremov+elmegreen98} find a correlation between the separation of stars and star
clusters and their age, and interpret that result in terms of the linewidth-size
relation in the ISM first discussed by \cite{larson81}. As velocity and size converts
into a timescale, they conclude that the formation of molecular clouds and stellar birth in
their interior proceeds on timescales of a few turbulent crossing times \citep{elmegreen00,
elmegreen07} \citep[see also][]{ballesteros-paredes99, hartmann01, glover10} 
which is a signpost of star formation controlled by interstellar
turbulence \citep{elmegreen+efremov97, maclowklessen04}. Further evidence is
provided by the fact that the mass spectrum of molecular clouds and clumps within
the clouds follows a power law, $dN/dM \propto M^{-\alpha}$, with an exponent in the
range $1.5 \lesssim \alpha \lesssim 2.0$ \citep{stutzki90, williams94, kramer98} 
which is very similar to the mass spectra of young
star clusters as discussed in \S~\ref{s:cmf}. The value of $\alpha = 1.5 \pm 0.7$ that we
derive for the young clusters in NGC~6822 is fully consistent with this picture. 

As the ISM exhibits a very complex, hierarchical morphological structure, it stands to
reason that stars follow a similar spatial pattern. We have studied that
aspect in \S~\ref{s:methstruct} in terms of the distributions of blue stars in individual 
stellar structures. Using the 10th nearest neighbor map to identify clusters, we
find that indeed these stars in NGC~6822 can be grouped into larger and larger
aggregates in a hierarchical fashion when we vary the detection threshold. Similar
results are found in the Magellanic Clouds (\citealp{gieles08, bastian09}; see also 
\citealp{hunter03}), M33 \citep{bastian07}, and M51 \citep{bastian05}. 
We note that the \Q-parameters derived for the individual
star clusters in NGC~6822 indicate that the stars in the clusters themselves are hierarchically
structured. It is a common feature of young star clusters that they reveal a high
degree of sub-clustering when observed with sufficient resolution and sensitivity to
identify individual stars down to the peak of the IMF \citep[e.g.,][]{cw04, cw08, cw09, 
schmeja08, schmeja09} with values that are
consistent with numerical calculations of star-cluster formation \citep{schmeja04}. 
We conclude that we find direct evidence, that the blue stars in NGC~6822 exhibit
a hierarchical spatial pattern from the scales of individual objects all the way up
to the scale of the galaxy as a whole. 

It is interesting to note in this context that when it is indeed true that the
spatial distribution of stars and star clusters traces the statistical properties of
ISM turbulence, then the absence of a clear break in the hierarchy could indicate
that turbulence is driven on kpc-scales. Possible mechanisms are global
gravitational instability \citep[e.g.,][]{li05}, the magneto-rotational instability
(MRI) in the disk of NGC~6822 \citep{beck96, balbus98, piontek07}, or 
the accretion of fresh gas through the halo in form of cold
streams or gas-rich satellites \citep{santillan07, sancisi08, klessen10}. 
Indeed, the "Northwestern Cloud" clearly visible in Fig.~\ref{f:maps}
has been speculated to be a separate system, possibly interacting with
NGC~6822 \citep{deblok00, deblok03}. This interaction
could also be responsible for a possible increase of the star formation rate in the
past 100 Myr (\citealp{hodge80}; \citealp[see however][for a different opinion]{gallart96a}).

\section{Summary}\label{s:sum}

In this paper we present a thorough investigation of the 
cluster population with age \lsim~500~Myr and the hierarchy 
in the spatial distribution of main-sequence stars of this 
age in the Local Group dwarf irregular galaxy NGC~6822. Our
observational material comprises optical imaging from the 
8.2-m {\sc Subaru} Telescope, providing the most complete
point source catalog of the galaxy in terms of dynamic range
and spatial coverage.

Star clusters are identified with the application of the 
Nearest-Neighbor (NN) density method for the 10th NN on
the blue main-sequence stars in NGC~6822. Fourty-seven 
distinct concentrations with $\geq$~3 stellar members that 
are found with our method to have 10th NN density values 
3$\sigma$ above the average background density are 
identified as the star clusters of the galaxy. The 
physical dimensions, stellar density and limiting 
radii of the clusters are defined by this density limit.
The size distribution of the detected clusters can be 
very well approximated by a Gaussian peaked at 
$\sim$~68~pc (with $\sigma \sim$~18~pc). 

The 
total stellar masses of the clusters are estimated 
by extrapolation of the total observed stellar mass function
of all clusters to the lower stellar masses, assuming
that it has the shape of the average Galactic  
mass function \citep{kroupa02} down to 0.1~M{\solar}. The derived cluster 
masses resign in the range $10^3$~-~$10^{4}$~M{\solar}.
Their distribution follows very well  a power-law 
with index $\sim -1.5 \pm 0.7$, somewhat shallower but consistent with 
the Cluster Initial Mass Function and the mass spectra found for clusters in other 
Local Group galaxies. Structural parameters, such 
as their Spitzer radius, the correspondent 3-dimensional 
stellar density and  velocity dispersion, as well as
their crossing and the two-body relaxation times
are computed for the clusters from their total masses
and stellar numbers. We find a correlation between the 
radii and stellar numbers of the detected clusters, 
which follows a power law similar to that found for 
Galactic embedded clusters. 

We study the clustering behavior of the blue stars in the considered
age-limit with the application of the NN density method and
the detection of stellar structures in different density levels,
other than 3$\sigma$ used for the establishment of the clusters.
We identify, thus, stellar concentrations of various length-scales,
clusters being the representative of one of them.  High-density 
stellar structures are revealed from the application of the 
NN method with higher (4$\sigma$ and 5$\sigma$) density 
threshold above the average density level, while the larger loose
stellar structures, which include the denser ones, are recognized
by the application of the method for lower (1$\sigma$ and 
2$\sigma$) density thresholds. All significant stellar concentrations, 
including the clusters, appear in the main (star-forming) part of the 
galaxy. The density maps constructed with this technique 
provide clear indications of hierarchically structured stellar 
concentrations in NGC~6822, in the sense that small dense 
stellar concentrations are located {\sl inside} larger and looser
ones. We illustrate this hierarchy by the so-called {\sl dendrogram}, or 
structure tree of the detected stellar structures in NGC~6822,
which demonstrates that most of the detected structures split up into 
several substructures over at least three levels. 

The majority of the detected star clusters in NGC~6822 has ages between 250 
and 500~Myr, which is longer than the crossing time of the galaxy, the timescale 
on which hierarchical cluster distributions in the LMC and SMC are erased 
\citep[e.g.,][]{bastian09}. As a consequence, the finding that these clusters have 
preserved their structure is in itself very interesting. However, our catalog of ``stellar 
clusters" comprises large stellar groups with sizes larger than the 
typical pc-scale Galactic clusters; the smallest cluster in our catalog is about 20 pc in size. 
As such, their large sizes is a possible explanation why  these clusters 
have preserved their structure longer than one crossing time of the galaxy.

We quantify the hierarchy of these structures with the use of the Minimum 
Spanning Tree method and the \Q\ parameter. We find that
the \Q\ values for the structures detected at 1, 2, and 3$\sigma$ 
density thresholds reside in the hierarchical regime 
($\Q < 0.8$), reaching \Q\ values that correspond to a fractal 
dimension of $D \approx 1.8$. The correlation between the 
number of stars and radius seen for the clusters of NGC~6822 
is also observed for the larger stellar structures, identified at the 
1$\sigma$ and 2$\sigma$ density levels, as well as for the 
smaller dense concentration identified at the 4$\sigma$ density 
level. It is worth noting that some of the large high-density 
stellar concentrations, particularly in the northern part of the 
star-forming portion of the galaxy, coincide with IR-bright structures
earlier identified with {\sl Spitzer}, associated with high column 
density ($\gsim~10^{21}~{\rm cm}^{-2}$) neutral gas, indicating 
structures that currently form stars.

The clusters of NGC~6822, which we identify at the 3$\sigma$ density level of 
the NN method, are found to be themselves internally structured. Considering 
the ages of these clusters, this result is also interesting, because as found 
for small compact clusters, any substructure is expected to have been rapidly 
erased on very short time-scales \citep{allison10}. However, the clusters we 
detect in NGC~6822 are large 
stellar groups, which may have survived strong disruptions for periods of time 
longer than their crossing times. Previous simulations of fractal star clusters 
by \cite{goodwinwhitworth04} have shown that even an initially 
homogeneous cluster can develop substructure, if it is born with coherent 
velocity dispersion. As a consequence, the substructure we observe in our 
clusters may not be the present exhibition of their original `primordial' 
substructure, but one induced later. The same authors find that the velocity 
dispersion is a key parameter, determining the survival of substructure, which 
can last for several crossing times. Therefore, it is possible that at least 
some of our clusters may have sustained their initial substructure. In favor to 
this, spatial substructure has been identified in open clusters of the Milky 
Way as old as $\sim$~100~Myr \citep{sanchez09}.

The morphological structure of the stellar concentrations we identified  
in NGC~6822 resembles the structural behavior of the interstellar 
matter, which in principal is {\sl hierarchical} and {\sl dominated by 
turbulent motions}. As such, this coincidence between the clustering 
of stars and the hierarchical structure of the ISM, as it is demonstrated in 
the present study, suggests that turbulence may play the major role in 
regulating clustered star formation in NGC~6822 on pc- and kpc-scales. 






%
%

\acknowledgements

D.A.G. acknowledges financial support from the German Research
Foundation (Deu\-tsche For\-schungs\-ge\-mein\-schaft, DFG) through 
grant GO~1659/1-2 and the German Aerospace Center (Deutsche Zentrum 
f\"{u}r Luft- und Raumfahrt, DLR) through grant 50~OR~0908. S.S.  
acknowledges the support of the German Research Foundation through 
grants SCHM 2490/1-1 and KL 1358/5-2.


\begin{thebibliography}{}

\bibitem[Adams et al.(2006)]{adams06}
Adams, F.~C., Proszkow, E.~M., Fatuzzo, M., \& Myers, P.~C.\ 2006, \apj, 641, 504

\bibitem[Allison et al.(2010)]{allison10} Allison, R.~J., 
Goodwin, S.~P., Parker, R.~J., Portegies Zwart, S.~F., 
\& de Grijs, R.\ 2010, \mnras, 1188 (arXiv:1004.5244)

\bibitem[Allen et al.(2007)]{allen07}
Allen, L., Megeath, S.~T., Gutermuth, R., et al. 2007, in Protostars and Planets~V, 
eds.\ B. Reipurth, D. Jewitt, \& K. Keil (Tucson: Univ.\ Arizona Press), 361

\bibitem[Balbus \& Hawley(1998)]{balbus98} 
Balbus, S.~A., \& Hawley, J.~F.\ 1998, Reviews of Modern Physics, 70, 1 

\bibitem[Ballesteros-Paredes et al.(1999)]{ballesteros-paredes99} 
Ballesteros-Paredes, J., Hartmann, L., \& V{\'a}zquez-Semadeni, E.\ 1999, \apj, 527, 285 

\bibitem[Ballesteros-Paredes et al.(2007)]{ballesteros-paredes07}
Ballesteros-Paredes, J., Klessen, R.~S., Mac Low, M.-M., \& Vazquez-Semadeni, E.\ 2007, Protostars and Planets V, 63 

\bibitem[Bastian et al.(2005)]{bastian05} 
Bastian, N., Gieles, M., Efremov, Y.~N., \& Lamers, H.~J.~G.~L.~M.\ 2005, \aap, 443, 79 

\bibitem[Bastian et al.(2007)]{bastian07} Bastian, N., Ercolano, 
B., Gieles, M., Rosolowsky, E., Scheepmaker, R.~A., Gutermuth, R., 
\& Efremov, Y.\ 2007, \mnras, 379, 1302 

\bibitem[Bastian et al.(2009)]{bastian09}
Bastian, N., Gieles, M., Ercolano, B., Gutermuth, R. 2009, \mnras, 392, 868

\bibitem[Bastian et al.(2010)]{bastian10} Bastian, N., Covey, 
K.~R., \& Meyer, M.~R.\ 2010, ARA\&A in press (arXiv:1001.2965) 

\bibitem[Battinelli et al.(2003)]{battinelli03} 
Battinelli, P., Demers, S., \& Letarte, B. 2003, A\&A, 405, 563

\bibitem[Beck et al.(1996)]{beck96} 
Beck, R., Brandenburg, A., Moss, D., Shukurov, A., \& Sokoloff, D.\ 1996, \araa, 34, 155 

\bibitem[Bik et~al.(2003)]{bik03}
{Bik} A.,  {Lamers} H.~J.~G.~L.~M.,  {Bastian} N.,  {Panagia} N.,
  {Romaniello} M.,  2003, \aap, 397, 473


\bibitem[Cannon et al.(2006)]{cannon06}
Cannon, J.~M., et al.\ 2006, \apj, 652, 1170

\bibitem[Carpenter(2000)]{carpenter00} 
Carpenter, J.~M.\ 2000, \aj, 120, 3139 

\bibitem[Cartwright \& Whitworth(2004)]{cw04}
Cartwright, A., \& Whitworth, A.~P. 2004, \mnras, 348, 589

\bibitem[Cartwright \& Whitworth(2008)]{cw08} 
Cartwright, A., \& Whitworth, A.~P.\ 2008, \mnras, 390, 807 

\bibitem[Cartwright \& Whitworth(2009)]{cw09}
Cartwright, A., \& Whitworth, A.~P.\ 2009, \mnras, 392, 341

\bibitem[Casertano \& Hut(1985)]{casertano85} 
Casertano, S., \& Hut, P.\ 1985, \apj, 298, 80 

\bibitem[Chabrier(2003)]{chabrier03} 
Chabrier, G.\ 2003, \pasp, 115, 763 


\bibitem[de Blok \& Walter(2000)]{deblok00} 
de Blok, W.~J.~G., \& Walter, F.\ 2000, \apjl, 537, L95 

\bibitem[de Blok \& Walter(2003)]{deblok03} 
de Blok, W.~J.~G., \& Walter, F.\ 2003, \mnras, 341, L39

\bibitem[de Blok \& Walter(2006)]{deblok06} 
de Blok, W.~J.~G., \& Walter, F.\ 2006, \aj, 131, 343 (dbW06)

\bibitem[de Grijs \& Anders(2006)]{degrijs06} 
de Grijs, R., \& Anders, P.\ 2006, \mnras, 366, 295

\bibitem[de Grijs \& Goodwin(2008)]{degrijs08}
{de Grijs} R.,  {Goodwin} S.~P.,  2008, \mnras, 383, 1000

\bibitem[de Grijs et~al.(2003)]{degrijs03}
{de Grijs} R.,  {Anders} P.,  {Bastian} N.,  {Lynds} R.,  {Lamers}
  H.~J.~G.~L.~M.,    {O'Neil} E.~J.,  2003, \mnras, 343, 1285


 \bibitem[Efremov et al.(1987)]{efremov87} 
 Efremov, I.~N., Ivanov, G.~R., \& Nikolov, N.~S.\ 1987, \apss, 135, 119 

\bibitem[Efremov \& Elmegreen(1998)]{efremov+elmegreen98}
Efremov, Y.~N., \& Elmegreen, B.~G. 1998, \mnras, 299, 588

\bibitem[Elmegreen(2000)]{elmegreen00} 
Elmegreen, B.~G.\ 2000, \apj, 530, 277 

\bibitem[Elmegreen(2007)]{elmegreen07} 
Elmegreen, B.~G.\ 2007, \apj, 668, 1064 

\bibitem[Elmegreen(2010)]{elmegreen10} 
Elmegreen, B.~G.\ 2010, IAU Symposium, 266,  Star clusters: basic galactic building 
blocks throughout time and space ed. R. de Grijs \& J. R. D. L\'{e}pine, (Cambridge: 
Cambridge Univ. Press), 3 

\bibitem[Elmegreen \& Falgarone(1996)]{elmegreenfalgarone96} 
Elmegreen, B.~G., \& Falgarone, E.\ 1996, \apj, 471, 816 

\bibitem[Elmegreen \& Efremov(1997)]{elmegreen+efremov97} 
Elmegreen, B.~G., \& Efremov, Y.~N.\ 1997, \apj, 480, 235 

\bibitem[Elmegreen \& Scalo(2004)]{elmegreenscalo04} 
Elmegreen, B.~G., \& Scalo, J.\ 2004, \araa, 42, 211 

\bibitem[Elmegreen et al.(2000)]{elmegreen00}
Elmegreen, B.~G., Efremov, Y., Pudritz, R.~E., \& Zinnecker, H. 2000, 
in Protostars and Planets~IV, eds.\ V. Mannings, A.~P. Boss, \& S.~S. Russell
(Tucson: Univ.\ Arizona Press), p.~179


\bibitem[Federrath et al.(2009)]{federrath09} 
Federrath, C., Klessen, R.~S., \& Schmidt, W.\ 2009, \apj, 692, 364 

\bibitem[Federrath et al.(2010)]{federrath10} 
Federrath, C., Roman-Duval, J., Klessen, R.~S., Schmidt, W., \& Mac Low, M.-M.\ 2010, \aap, 512, A81 

\bibitem[Gallart et al.(1996a)]{gallart96a} 
Gallart, C., Aparicio, A.,  Bertelli, G., \& Chiosi, C.\ 1996a, \aj, 112, 1950

\bibitem[Gallart et al.(1996b)]{gallart96b}
Gallart, C., Aparicio, A., Bertelli, G., \& Chiosi, C. 1996b, AJ, 112, 2596

\bibitem[Gieles et al.(2008)]{gieles08} 
Gieles, M., Bastian, N., \& Ercolano, B.\ 2008, \mnras, 391, L93 

\bibitem[Gieles(2009)]{gieles09} Gieles, M.\ 2009, \mnras, 394, 
2113 

\bibitem[Girardi et al.(2002)]{girardi02} 
Girardi, L., Bertelli, G., Bressan, A., Chiosi, C., Groenewegen, M.~A.~T., Marigo, P., Salasnich, B., \& Weiss, A.\ 2002, \aap, 391, 195 

\bibitem[Glover et al.(2010)]{glover10} 
Glover, S.~C.~O., Federrath, C., Mac Low, M.-M., \& Klessen, R.~S.\ 2010, \mnras, 404, 2 

\bibitem[Goodwin \& Whitworth(2004)]{goodwinwhitworth04} 
Goodwin, S.~P., \& Whitworth, A.~P.\ 2004, \aap, 413, 929

\bibitem[Gouliermis et al.(2002)]{gouliermis02} Gouliermis, D., Keller,
S.~C., de Boer, K.~S., Kontizas, M., \& Kontizas, E.\ 2002, \aap, 381,
862

\bibitem[Gouliermis et al.(2003)]{gouliermis03} 
Gouliermis, D., Kontizas, M., Kontizas, E., \& Korakitis, R.\ 2003, \aap, 405, 111 

\bibitem[Gouliermis(2010)]{gouliermis09} 
Gouliermis, D. A.,\ 2010, in  {\sl  Star-forming dwarf galaxies:  
following Ariadne's thread in the cosmic labyrinth}, Physica Scripta, in press

\bibitem[Hartmann et al.(2001)]{hartmann01} 
Hartmann, L., Ballesteros-Paredes, J., \& Bergin, E.~A.\ 2001, \apj, 562, 852 

\bibitem[Heitsch et al.(2001)]{heitsch01} 
Heitsch, F., Mac Low, M.-M., \& Klessen, R.~S.\ 2001, \apj, 547, 280 

\bibitem[Hennebelle \& Chabrier(2008)]{hennebelle08} 
Hennebelle, P., \& Chabrier, G.\ 2008, \apj, 684, 395 

\bibitem[Hennebelle \& Chabrier(2009)]{hennebelle09} 
Hennebelle, P., \& Chabrier, G.\ 2009, \apj, 702, 1428 

\bibitem[Hodge(1980)]{hodge80} 
Hodge, P.W., 1980, \apj, 241, 125


\bibitem[Houlahan \& Scalo(1992)]{houlahan92}
Houlahan, P., \& Scalo, J.\ 1992, \apj, 393, 172

\bibitem[Hunter \& Elmegreen(2004)]{hunter04} 
Hunter, D.~A., \& Elmegreen, B.~G.\ 2004, \aj, 128, 2170 

\bibitem[Hunter et al.(2003)]{hunter03} 
Hunter, D.~A., Elmegreen, B.~G., Dupuy, T.~J., Mortonson, M.\
2003, AJ, 126, 1836 

\bibitem[Israel et al.(1996)]{israel96} 
Israel, F.P., Bontekoe, Tj.R., Kester, D.J.M, 1996, \aap, 308, 723

\bibitem[Ivanov(1996)]{ivanov96}
Ivanov, G. R. 1996, \aap, 305, 708

\bibitem[Karampelas et al.(2009)]{karampelas09} 
Karampelas, A., Dapergolas, A., Kontizas, E., Livanou, E., Kontizas, M., Bellas-Velidis, I., \& V{\'{\i}}lchez, J.~M.\ 2009, \aap, 497, 703 

\bibitem[Klessen(2001a)]{klessen01a} 
Klessen, R.~S.\ 2001a, \apj, 556, 837 

\bibitem[Klessen(2001b)]{klessen01b} 
Klessen, R.~S.\ 2001b, \apjl, 550, L77 

\bibitem[Klessen \& Burkert(2000)]{klessenburkert00} 
Klessen, R.~S., \& Burkert, A.\ 2000, \apjs, 128, 287 

\bibitem[Klessen \& Burkert(2001)]{klessenburkert01} 
Klessen, R.~S., \& Burkert, A.\ 2001, \apj, 549, 386 

\bibitem[Klessen \& Hennebelle(2010)]{klessen10} 
Klessen, R.~S., \& Hennebelle, P.\ 2010, A\&A in press (arXiv:0912.0288)

\bibitem[Klessen et al.(2000)]{klessenetal00} 
Klessen, R.~S., Heitsch, F., \& Mac Low, M.-M.\ 2000, \apj, 535, 887 

\bibitem[Komiyama et al. (2003)]{komiyama03}
Komiyama, Y., et al. 2003, \apj, 590, L17

\bibitem[Kramer et al.(1998)]{kramer98} 
Kramer, C., Alves, J., Lada, C., Lada, E., Sievers, A., Ungerechts, H., \& Walmsley, M.\ 1998, \aap, 329, L33 

\bibitem[Kroupa(2002)]{kroupa02} 
Kroupa, P.\ 2002, Science, 295, 82

\bibitem[Kroupa(2008)]{kroupa08} 
Kroupa, P.\ 2008, Lecture Notes in Physics, Berlin Springer Verlag, 760, 181 

\bibitem[Krumholz \& McKee(2005)]{krumholz05} 
Krumholz, M.~R., \& McKee, C.~F.\ 2005, \apj, 630, 250 

\bibitem[Lada \& Lada (2003)]{lada03} Lada, C. J., \& Lada, E. A. 2003,
ARA\&A, 41, 57


\bibitem[Larson(1981)]{larson81}
Larson, R.~B. 1981, \mnras, 194, 809

\bibitem[Li et al.(2005)]{li05} 
Li, Y., Mac Low, M.-M., \& Klessen, R.~S.\ 2005, \apjl, 620, L19 

\bibitem[Mac Low \& Klessen(2004)]{maclowklessen04} 
Mac Low, M.-M., \& Klessen, R.~S.\ 2004, Reviews of Modern Physics, 76, 125 

\bibitem[Ma\'{i}z Apell\'{a}niz \& \'{U}beda(2005)]{maizapellaniz05} 
Ma\'{i}z Apell\'{a}niz, J., \& \'{U}beda, L. 2005, ApJ, 629, 873 
 
\bibitem[Massey(2006)]{massey06a} 
Massey, P. 2006, The Local Group as an Astrophysical Laboratory, 164

\bibitem[Massey et al.(2007)]{massey07} 
Massey, P., Olsen, K.~A.~G., Hodge, P.~W., Jacoby, G.~H., McNeill, R.~T., Smith, R.~C., 
\& Strong, S.~B.\ 2007, \aj, 133, 2393 

\bibitem[Massey et al.(2006)]{massey06} 
Massey, P., Olsen, K.~A.~G., Hodge, P.~W., Strong, S.~B., Jacoby, G.~H., Schlingman, W., 
\& Smith, R.~C.\ 2006, \aj, 131, 2478 

\bibitem[Mateo(1998)]{mateo98} 
Mateo, M.L. 1998, \araa, 36, 435

\bibitem[McCrady \& Graham(2007)]{mccrady07}
{McCrady} N.,  {Graham} J.~R.,  2007, \apj, 663, 844

\bibitem[Miller \& Scalo (1979)]{miller79} 
Miller, G.~E., \& Scalo, J.~M.\ 1979, ApJS, 41, 513

\bibitem[Miyazaki et al.(2002)]{miyazaki02} 
Miyazaki, S., et al.\  2002, \pasj, 54, 833 

\bibitem[Piontek \& Ostriker(2007)]{piontek07} 
Piontek, R.~A., \& Ostriker, E.~C.\ 2007, \apj, 663, 183 

\bibitem[Rom{\'a}n-Z{\'u}{\~n}iga et al.(2008)]{romanzuniga08} 
Rom{\'a}n-Z{\'u}{\~n}iga, C.~G., Elston, R., Ferreira, B., \& Lada, E.~A.\ 2008, \apj, 672, 861 

\bibitem[Rosolowsky et al.(2008)]{rosolowsky08}
Rosolowsky, E.~W., Pineda, J.~E., Kauffmann, J., \& Goodman, A.~A.\ 2008, \apj, 679, 1338

\bibitem[Salpeter(1955)]{salpeter55} 
Salpeter, E. E. 1955, ApJ, 121, 161

\bibitem[S{\'a}nchez \& Alfaro(2009)]{sanchez09} 
S{\'a}nchez, N., \& Alfaro, E.~J.\ 2009, \apj, 696, 2086

\bibitem[Sancisi et al.(2008)]{sancisi08} 
Sancisi, R., Fraternali, F., Oosterloo, T., \& van der Hulst, T.\ 2008, \aapr, 15, 189 

\bibitem[Santill{\'a}n et al.(2007)]{santillan07} 
Santill{\'a}n, A., S{\'a}nchez-Salcedo, F.~J., \& Franco, J.\ 2007, \apjl, 662, L19 

\bibitem[Scalo(1986)]{scalo86} 
Scalo, J.\ 1986, Fundam. Cosmic Phys., 11, 1

\bibitem[Scalo(1998)]{scalo98}
Scalo, J. 1998, in ASP Conf. Ser. 142, The Stellar Initial Mass Function (38th 
Herstmonceux Conf.), ed. G. Gilmore \& D. Howell (San Francisco, CA: 
ASP), 201

\bibitem[Scalo \& Elmegreen(2004)]{scaloelmegreen04} 
Scalo, J., \& Elmegreen, B.~G.\ 2004, \araa, 42, 275 

\bibitem[Schmeja(2010)]{schmeja10} 
Schmeja, S.,\ 2010, Astron. Nachr. accepted

\bibitem[Schmeja \& Klessen(2004)]{schmeja04} 
Schmeja, S., \& Klessen, R.~S.\ 2004, \aap, 419, 405 

\bibitem[Schmeja \& Klessen(2006)2006]{sk06}
Schmeja, S., \& Klessen, R.~S. 2006, \aap, 449, 151

\bibitem[Schmeja et al.(2008)]{schmeja08} 
Schmeja, S., Kumar, M.~S.~N., \& Ferreira, B.\ 2008, \mnras, 389, 1209 

\bibitem[Schmeja et al.(2009)]{schmeja09} 
Schmeja, S., Gouliermis, D.~A., \& Klessen, R.~S.\ 2009, \apj, 694, 367 


\bibitem[Skillman et al.(1989)]{skillman89} 
Skillman, E.D., Terlevich, R., Melnick, J. 1989, \mnras, 240, 563

\bibitem[Spitzer(1987)]{spitzer87} 
Spitzer, L.\ 1987, Princeton, NJ, Princeton University Press, 1987, 191 p.

\bibitem[Spitzer(1958)]{spitzer58} 
Spitzer, L., Jr.\ 1958, \apj, 127, 17

\bibitem[Spitzer(1969)]{spitzer69} 
Spitzer, L., Jr. 1969, ApJ, 158, L139 

\bibitem[Stutzki \& Guesten(1990)]{stutzki90} 
Stutzki, J., \& Guesten, R.\ 1990, \apj, 356, 513 

\bibitem[Stutzki et al.(1998)]{stutzki98} 
Stutzki, J., Bensch, F., Heithausen, A., Ossenkopf, V., \& Zielinsky, M.\ 1998, \aap, 336, 697 

\bibitem[van den Bergh(2000)]{vandenbergh00} 
van den Bergh, S.\ 2000, Cambridge Astrophysics Series, 35,  

\bibitem[V{\'a}zquez-Semadeni et al.(2003)]{vazquez-semadeni03} 
V{\'a}zquez-Semadeni, E., Ballesteros-Paredes, J., \& Klessen, R.~S.\ 2003, \apjl, 585, L131 

\bibitem[V{\'a}zquez-Semadeni et al.(2009)]{vazquez-semadeni09} 
V{\'a}zquez-Semadeni, E., G{\'o}mez, G.~C., Jappsen, A.-K., Ballesteros-Paredes, J., \& Klessen, R.~S.\ 2009, \apj, 707, 1023 

\bibitem[von Hoerner(1963)]{vonhoerner63}
von Hoerner, S. 1963, \zap, 57, 47

\bibitem[Weldrake et al.(2003)]{weldrake03}
Weldrake, D. T. F., de Blok, W. J. G., \& Walter, F. 2003, MNRAS, 340, 12

\bibitem[Weisz et al.(2008)]{weisz08} Weisz, D.~R., Skillman, 
E.~D., Cannon, J.~M., Dolphin, A.~E., Kennicutt, R.~C., Jr., Lee, J., 
\& Walter, F.\ 2008, \apj, 689, 160 


\bibitem[Williams et al.(1994)]{williams94} 
Williams, J.~P., de Geus, E.~J., \& Blitz, L.\ 1994, \apj, 428, 693 

\bibitem[Wyder (2001)]{wyder01} 
Wyder, T. K. 2001, AJ, 122, 2490

\bibitem[Zhang \& Fall(1999)]{zhang99}
{Zhang} Q.,  {Fall} S.~M.,  1999, \apjl, 527, L81

\end{thebibliography}
\end{document}